\documentclass[]{pasj01_revised}

\begin{document} 

\title{Japanese Cosmic Dawn/Epoch of Reionization Science with the Square Kilometre Array}
\author{Kenji \textsc{Hasegawa}\altaffilmark{1*}, 
Shinsuke \textsc{Asaba}\altaffilmark{1}, 
Kiyotomo \textsc{Ichiki}\altaffilmark{1}, 
Akio K. \textsc{Inoue}\altaffilmark{2}, 
Susumu \textsc{Inoue}\altaffilmark{3}, 
Tomoaki \textsc{Ishiyama}\altaffilmark{4}, 
Hayato \textsc{Shimabukuro}\altaffilmark{1,5}, 
Keitaro \textsc{Takahashi}\altaffilmark{5}, 
Hiroyuki \textsc{Tashiro}\altaffilmark{1}, 
Hidenobu \textsc{Yajima}\altaffilmark{6}, 
Shu-ichiro \textsc{Yokoyama}\altaffilmark{7}, 
Kohji \textsc{Yoshikawa}\altaffilmark{8}, 
Shintaro \textsc{Yoshiura}\altaffilmark{5}, 
on behalf of Japan SKA Consortium (SKA-JP) EoR Science Working Group}
\altaffiltext{1}{Department of Physics and Astrophysics, Nagoya University Furo-cho, Chikusa-ku, Nagoya, Aichi 464-8602, Japan}
\altaffiltext{2}{College of General Education, Osaka Sangyo University, 3-1-1, Nakagaito, Daito, Osaka 574-8530, Japan}
\altaffiltext{3}{Astrophysical Big Bang Laboratory, Riken, Wako, Saitama 351-0198, Japan}
\altaffiltext{4}{Institute of Management and Information Technologies, Chiba University, 1-33, Yayoi-cho, Inage-ku, Chiba, 263-8522, Japan}
\altaffiltext{5}{Faculty of Science, Kumamoto University, 2-39-1 Kurokami, Kumamoto 860-8555, Japan}
\altaffiltext{6}{Frontier Research Institute for Interdisciplinary Sciences, Tohoku University, Sendai 980-8578, Japan}
\altaffiltext{7}{Department of Physics, Rikkyo University, 3-34-1 Nishi-Ikebukuro, Toshima, Tokyo
171-8501, Japan}
\altaffiltext{8}{Center for Computational Sciences, University of Tsukuba, 1-1-1, Ten-nodai, Tsukuba, Ibaraki, 305-8577, Japan}
\email{hasegawa.kenji@a.mbox.nagoya-u.ac.jp}

\KeyWords{reionization, first stars, galaxies} 

\maketitle
\begin{figure}
\vspace{-233mm}
\includegraphics[width=25mm]{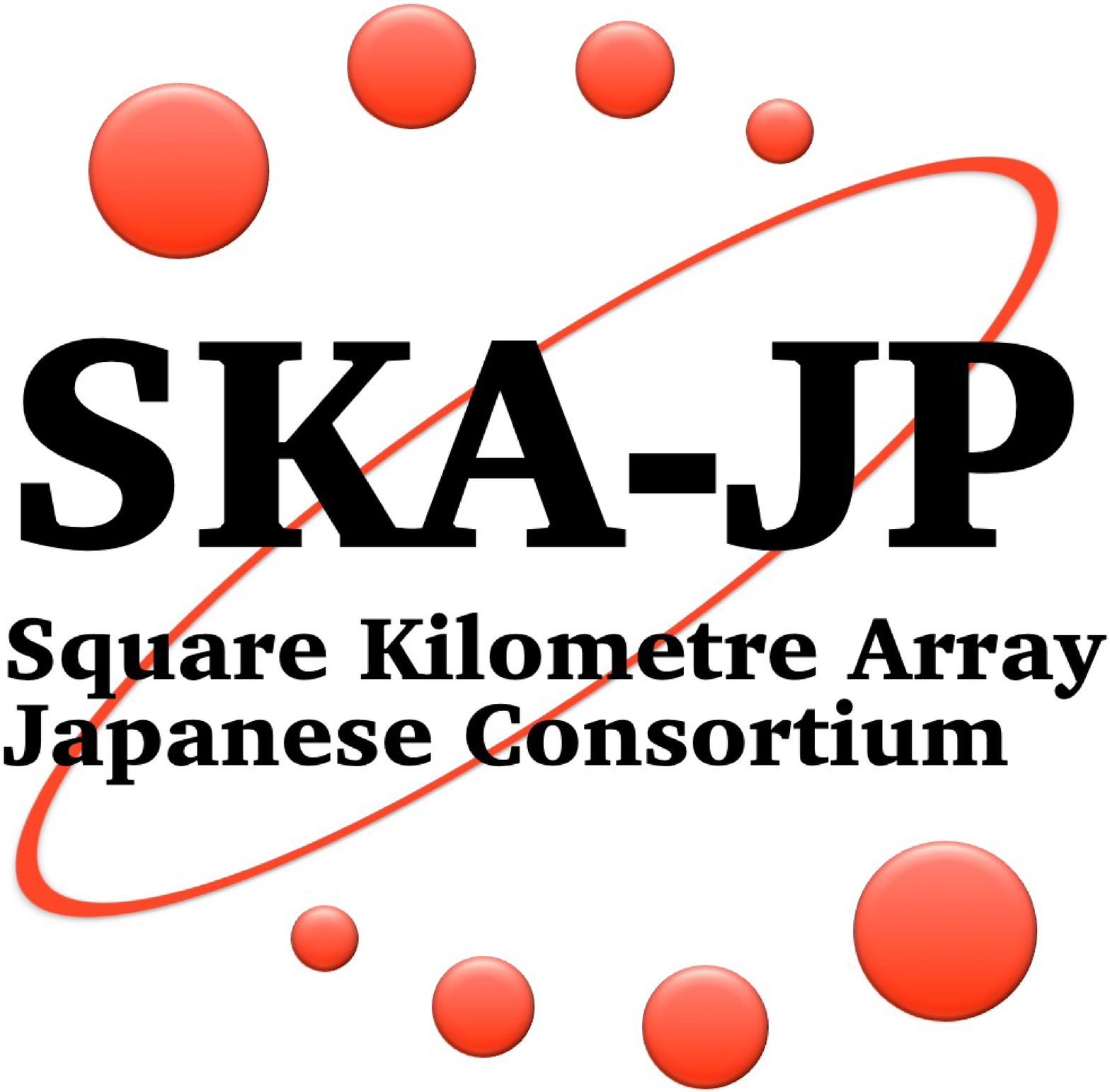}
\vspace{177mm}
\end{figure}
\begin{abstract}
Cosmic reionization is known to be a major phase transition of the gas in the Universe. 
Since astronomical objects formed in the early Universe, such as the first stars, galaxies and black holes, are expected to have caused cosmic reionization, the formation history and properties of such objects are closely related to the reionization process.  
In spite of the importance of exploring reionization, our understandings regarding reionization is not sufficient yet. 
Square Kilometre Array (SKA) is a next-generation large telescope that will be operated in the next decade.  
Although several programs of next-generation telescopes are currently scheduled, the SKA will be the unique telescope with a potential to directly observe neutral hydrogen up to $z\approx30$, and provide us with valuable information on the Cosmic Dawn (CD) and the Epoch of Reionization (EoR). 
The early science with the SKA will start in a few years; it is thus the time for us to elaborate a strategy for CD/EoR Science with the SKA. 
The purpose of this document is to introduce Japanese scientific interests in the SKA project and to report results of our investigation. 
\end{abstract}
\section{Introduction}
The epoch shortly after cosmic recombination is often called Dark Ages (DA), since there were no luminous objects at the epoch. 
Recent theoretical studies have predicted that the formation of luminous first stars marked the end of the DA (e.g., \cite{2002Sci...295...93A}, \cite{2002ApJ...564...23B}, \cite{2006ApJ...652....6Y}). 
The first stars are expected to have firstly illuminated the Universe, hence this very early phase of the structure formation is called Cosmic Dawn (CD). 
According to the standard hierarchical structure formation scenario, more massive objects, such as galaxies, were formed following the CD, and the increase in the number of ionizing photon sources would result in the ionization and heating of the Universe. 
This global transition is known as cosmic reionization, and the epoch when reionization dramatically proceeded is the Epoch of Reionization (EoR). 
Recent observations have provided us with fruitful information on cosmic reionization. 
For instance the spectra of high-$z$ quasars (QSOs) tell us that  the Universe was almost ionized by $z\sim6$ (e.g., \cite{2006ARA&A..44..415F}), and the decrease in the number density of Ly$\alpha$ emitters (LAEs) from $z\sim6$ to $z\sim7$ implies that neutral hydrogen fraction increased with redshift at $z>6$ (e.g., \cite{2014ApJ...797...16K}). 
In addition, the optical depth to Thomson scattering of Cosmic Microwave Background (CMB) photons measured by the Planck satellite is $\tau_{\rm e}=0.066\pm0.016$, which corresponds to an instantaneous reionization redshift of $z_{\rm r}=8.8^{+1.7}_{-1.4}$ \citep{2015arXiv150201589P}. 
Therefore a gradual reionization history is required for satisfying all of the observational facts. 
In contrast to the advancing knowledges on the time evolution of cosmic reionization, it is still difficult to impose critical constraints on the types of ionizing sources and the topology of reionization, because we have never had key observations to know the spatial distribution of neutral hydrogen during the EoR.  

The {\it Square Kilometre Array} (SKA) is a next-generation telescope that will be fully operated in the next decade, and is expected to be capable of detecting the redshifted hyper-fine structure transition line of neutral hydrogen, i.e. 21-cm line, during the CD/EoR. 
Therefore it will be possible to know more detailed information on neutral hydrogen up to $z\approx30$ with the SKA. 
In the international SKA science book published in 2015 \footnote{The international SKA science book can be downloaded from https://www.skatelescope.org/books/}, the international SKA CD/EoR 
science working group (SWG) members have proposed what we will be able to learn from observations with the SKA. 
Although some studies on the CD/EoR by Japanese scientists are already described in the international SKA science book, there are additional activities that significantly contribute to the CD/EoR Science with the SKA. 
Hence the purpose of this document is to introduce Japanese scientific interests in the SKA project and to report results of our investigation. It is still in progress, so that the document may not fully cover previous works related to the SKA. We wish that the document becomes an interface for future communications, collaborations, and synergies with worldwide communities. 

This document is organized as follows. 
In section 2, we briefly review some topics on the CD/EoR proposed in the international SKA science book. 
Subsequently in section 3, the main body of this article, we introduce the current status and future prospects of SKA-Japan consortium (SKA-JP) EoR SWG. 
Finally section 4 is devoted to the summary. 

\section{Current status of CD/EoR Science with the SKA}\label{SB}
For the purpose of achieving various CD/EoR science goals, three key observational plans, the shallow, medium, and deep surveys, are currently proposed for SKA1-low.
The survey area and the observing time are 10,000 sq. degree and 10 hours/field for the shallow survey, 1,000 sq. degree and 100 hours/field for the medium survey, and 100 sq. degree and 1,000 hours/field for the deep survey. 
During the EoR, 1 degree roughly corresponds to $\sim 100h^{-1}{\rm Mpc}$. 
The shallow and medium surveys will be carried out mainly for the power spectrum analysis, and the deep survey mainly for the imaging analysis. 
As for the 21-cm forest analysis, a different kind of observation in which 1,000 hours integration targeting a high-$z$ radio loud source will be required. 
In the international SKA science book published in 2015 (e.g., \cite{2015aska.confE...1K}; \cite{2015aska.confE...3A}; \cite{2015aska.confE...6C}; \cite{2015aska.confE...8J}; \cite{2015aska.confE...9M}; \cite{2015aska.confE..10M}; \cite{2015aska.confE..11M}; \cite{2015aska.confE..12P}; \cite{2015aska.confE..13S}; \cite{2015aska.confE..15W}), numbers of topics on the CD/EoR science have already reported. 
In this section, we briefly introduce major CD/EoR Science partly developed by the international SKA CD/EoR SWG, and partly presented in the international SKA Science Book.

\subsection{Basics of 21-cm signals during CD/EoR}
When we observe redshifted HI 21-cm signals during the EoR, the signals will be observed as the difference from the CMB light. 
Previous studies (e.g., \cite{1997ApJ...475..429M}) have shown that the deferential brightness temperature $\delta T_{\rm b}$ can be written as
\begin{eqnarray}
	\delta T_{\rm b}  &\approx& 27 x_{\rm HI} (1+\delta) \left( \frac{1+z}{10}\right)^{\frac{1}{2}} \left( 1-\frac{T_{\rm CMB}(z)}{T_{\rm s}}\right) \nonumber \\	&& \times \left( \frac{\Omega_{\rm b}}{0.044}\frac{h}{0.7}\right) \left( \frac{\Omega_{\rm m}}{0.27}\right)^{\frac{1}{2}} \left( \frac{1-Y_{\rm p}}{1-0.248}\right) \nonumber \\ && \times \left( 1+\frac{1}{H(z)}\frac{dv_\parallel}{dr_\parallel}\right)^{-1} [\rm mK], \label{dts}
\end{eqnarray}
where $\Omega_{\rm b}$, $\Omega_{\rm m}$, $h$, $H(z)$ and $Y_{\rm p}$ are cosmological parameters as they usually means. In addition, $\delta \equiv \rho/\langle \rho \rangle -1$, $x_{\rm HI}$, $T_{\rm CMB}$ and $T_{\rm s}$ are the baryonic overdensity, the neutral hydrogen fraction, the CMB temperature and the spin temperature of the 21-cm transition, respectively. 
Eq.~(\ref{dts}) indicates that the brightness temperature $\delta T_{\rm b}$ in a warm ($T_{\rm s}>T_{\rm CMB}$) region is positive, while that in a cold ($T_{\rm s}<T_{\rm CMB}$) region is negative.  
In the particular case of $T_{\rm s} \gg T_{\rm CMB}$ the brightness temperature is independent of the spin temperature anymore. 
As shown by Eq.~(\ref{dts}), the ionization state of the intergalactic matter (IGM) is important for determining the amplitude of $\delta T_{\rm b}$.
It is also important to know the thermal state of the IGM, since the spin temperature $T_{\rm s}$ is expected to be coupled with gas temperature $T_{\rm gas}$ via atomic collisions and Ly$\alpha$ pumping (Wouthuysen-FIeld (WF) effect, \cite{1952AJ.....57R..31W}, \cite{1959ApJ...129..536F}). 
The ionization and thermal states of the IGM are essentially determined by properties of ionizing sources, such as emissivity, spectral type and spatial distribution. 
Often the properties of ionizing sources are regulated by some astrophysical processes called feedback that will be described in the next section.  

\subsection{Ionizing sources and related feedback processes}\label{physics}
Observationally  there is no crucial evidence for the kinds of the objects that actually caused cosmic reionization, but some candidates have been proposed theoretically. 
The first stars were presumably the first ionizing sources, because they are thought to have been typically massive and emitting sufficient ionizing photons \citep{2002Sci...295...93A, 2002ApJ...564...23B, 2008Sci...321..669Y, 2011Sci...334.1250H, 2014ApJ...781...60H, 2014ApJ...792...32S}. 
It is important to consider feedback effects for quantifying the contribution of the first stars to cosmic reionization as discussed later. 
After the epoch of the first star formation, high-$z$ galaxies are plausible candidates of ionizing sources, since a large number of galaxies have been already discovered at $z>6$ \citep{2010ApJ...723..869O, 2014ApJ...797...16K, 2015ApJ...803...34B, 2014ApJ...786..108O, 2015ApJ...802L..19R}. 
One of the important processes deciding the number of ionizing photons provided by the galaxies is the escape of ionizing continuum radiation. 
Many observational and theoretical studies have proposed various values of the escape fraction $f_{\rm esc}$, hence there is no obvious consensus on the value of $f_{\rm esc}$, although most of numerical studies have shown a similar trend, i.e. $f_{\rm esc}$ decreases with the halo mass \citep{2010ApJ...710.1239R, 2011MNRAS.412..411Y,2014MNRAS.442.2560W,2015MNRAS.451.2544P}. 
Feedback effects are also important, since they play important roles in deciding the intrinsic ionizing continuum emissivity by regulating star formation rates (SFRs) in galaxies.
It is also expected that X-ray radiation from X-ray binaries and active galactic nuclei (AGNs) contributed to the ionization and heating of the Universe. 
As shown later in \S\ref{PS} and \S\ref{imaging}, the impact by high energy X-ray photons on HI 21-cm signals are expected to be significantly different from those by ultraviolet (UV) photons because of long mean free paths of the high energy photons. 

As mentioned above, feedback effects severely affect the number of ionizing photons from the ionizing sources. 
It is widely known that radiative feedback is one of important feedback processes. 
Since ionizing photons heat gas temperature up to $\sim 10^4$K, the photo-heating leads to the evaporation of low-mass systems with virial temperatures of $T_{\rm vir} < 10^4$K \citep{2012MNRAS.422.3067P, 2012MNRAS.424..377D}. 
Even in more massive galaxies, photo-heating may decrease SFRs via preventing the interstellar gas from fragmentations \citep{ 2013MNRAS.428..154H}. 
\citet{2012MNRAS.427..311W} have shown that radiation pressure, in addition to the thermal pressure by the ionized gas, affects the motion of the interstellar gas. 
Expelling the gas from galaxies via the radiative \citep{2009ApJ...693..984W, 2012PTEP.2012aA306U} and supernova (SN) feedback \citep{2014ApJ...788..121K} also affects $f_{\rm esc}$. 
Photo-dissociation of hydrogen molecules $\rm H_2$, which are main coolant in primordial gas with $T_{\rm gas} < 10^4$~K, by Lyman-Werner (LW) radiation is another type of radiative feedback (e.g., \cite{2003ApJ...599..746O}; \cite{2007ApJ...659..908S}; \cite{2009MNRAS.395.1280H}). 
The long mean free path of LW radiation compared to ionizing radiation results in the global suppression of the first star formation \citep{2008ApJ...673...14O,2012ApJ...756L..16A,2015MNRAS.448..568H}.  

Recently \citet{2010PhRvD..82h3520T} have pointed out another type of feedback. 
They have shown that after cosmic recombination baryons and cold dark matter (CDM) had supersonic relative flow and that the relative motion can introduce non-negligible effect on the structure formation in the early Universe.
Compared with the sound speed of 6 km/s just after cosmic recombination, the rms of the relative velocity is around 30 km/s. 
Such supersonic relative velocity affects the formation of dark matter (DM) halos with the mass of $10^5$--$10^7$ $M_{\odot}$. 
Interestingly, DM halos with these mass scales are thought to have hosted the first generation of stars and galaxies in the cosmological context. 
Therefore, the relative velocity makes a sizable impact on the structure formation in the EoR and hence the progress of cosmic reionization \citep{2015aska.confE...9M}. 
Cosmological numerical simulations \citep{2012ApJ...747..128N,
2013ApJ...763...27N, 2011MNRAS.412L..40M} have shown significant decrease in the halo abundance and gas fraction in the DM halos under the presence of the supersonic relative velocity, which leads to the succeeding delay of the star formation activity and the reionization of the surrounding IGM. 

\subsection{Methods for Modeling Reionization}\label{modelling}
The redshifted 21-cm signals will provide full three-dimensional HI distributions during the CD/EoR. Then we need modeling for extracting what we want to know from the detected signals. For the purpose, direct numerical simulations and semi-numerical/analytic methods are currently popular ways. 

The approach of direct numerical simulations is further categorized into two broad types. 
The first type is radiation hydrodynamic (RHD) simulations resolving small-scale structures under the influence of the feedback mentioned in \S \ref{physics} \citep{1997ApJ...486..581G, 2000MNRAS.314..611C,2002ApJ...575...49R,2013MNRAS.428..154H}. 
The second type is large-scale simulations in which radiative transfer (RT) is usually dealt as post-processing \citep{2006MNRAS.369.1625I, 2007ApJ...671....1T, 2014MNRAS.439..725I}. 
The simulations belonging to the latter type allow us to assess the global reionization process and its observational signatures, but we usually need appropriate sub-grid models to consider the feedback and small-scale structures. 

The early approach of semi-numerical/analytic methods is mostly based on the excursion-set formalism in which the number of ionizing photons is estimated by using analytically derived halo mass functions \citep{2004ApJ...613...16F, 2005MNRAS.363..818S,2006ApJ...653..815M}. 
The advantages of the semi-numerical approach compared to direct RT simulations are the speed and ease of use. 
Recent semi-numerical simulations take spatial distributions into account by adopting Lagrangian Perturbation Theory (LPT) or $N$-body simulation \citep{2007ApJ...654...12Z, 2009MNRAS.394..960C, 2011MNRAS.411..955M, 2013MNRAS.428.2467K}, but the evolution of HII bubbles is simply evaluated by the balance between the number of produced ionizing photons and that of atoms, instead of solving RT. 
Results by semi-numerical approaches often show good agreement with those by RT simulations on large-scales ($\gtrsim$Mpc) (e.g., \cite{2014MNRAS.443.2843M}), however further comparisons including detailed physics would be required. 

\subsection{Power-spectrum analysis}\label{PS}
In HI 21-cm power spectrum analysis, we often focus on three remarkable epochs \citep{2006MNRAS.371..867F, 2014MNRAS.439.3262M, 2015aska.confE...3A}. 
Firstly, redshifted UV radiation emitted by early stellar population is expected to contribute to Ly$\alpha$ pumping, and induce a remarkable peak on the HI 21-cm power-spectrum. 
It is often believed that  X-ray heating epoch follows the Ly$\alpha$ pumping epoch. 
As already mentioned in \S\ref{physics}, much longer mean free paths of X-ray photons lead to mostly uniform heating of the IGM and produce another peak in the large-scale power spectrum. 
Finally in the EoR, the power spectrum is again boosted due to the quite inhomogeneous HI distribution. 

Observation of the HI 21-cm signals is a big technical challenge because we have a number of possible contaminations including foreground emission by the Galaxy and nearby radio galaxies, and the ionospheric emission of the earth atmosphere. Furthermore, the intensity of the HI 21-cm signals is much smaller than those of contamination sources by many orders of magnitude, and its detection is technically quite challenging. 
Cross-correlation with data in different wavebands is an effective way to detect signals from noisy observational data.
Since the CMB data contain the kinematic Sunyaev--Zel'dovich (kSZ) effect which is the temperature fluctuation originated from the Thomson scattering during the EoR, cross-correlation with the CMB data could yield interesting information on the EoR.  
Many theoretical and numerical studies \citep{2006ApJ...647..840A,2008MNRAS.384..291A,2004PhRvD..70f3509C,2007MNRAS.377..168S,2010MNRAS.402.2617T} have shown that the kSZ effect and the HI 21-cm signals anti-correlate on the scales of the typical size of ionized bubbles, and correlate on larger scales. 
In particular, the correlation on larger scales depends on the rate of overall reionization \citep{2010MNRAS.402.2617T}. 

Correlation with high redshift galaxies is also important because they are tracers of cosmological density field as well as the sources of cosmic reionization \citep{2009ApJ...690..252L,2013MNRAS.432.2615W}.
Before and in the early phase of cosmic reionization, the HI 21-cm signals and the galaxy number density are expected to correlate on large scales. In the intermediate stage of reionization, they anti-correlate because ionizing photons emitted by high redshift galaxies ionize the surrounding IGM and hence reduce the HI 21-cm signals, where the scales of the anti-correlation reflect the size of the ionized bubbles. 

It is also expected that significant fraction of galaxies and stars in the EoR can be observed in the form of diffuse near infrared radiation. 
Thus, in the same manner, we can expect the cross-correlation of the HI 21-cm signals with the the near infrared background (NIRB).
\citet{2014MNRAS.440..298F} have investigated the HI 21-cm signals and NIRB based on the numerical simulations of cosmic reionization, and found that the NIRB anti-correlates with the HI 21-cm signals as reionization proceeds.

\subsection{Imaging analysis}\label{imaging}
When we observe a spatial distribution of the differential brightness temperature $\delta T_{\rm b}$ with a frequency $\nu_0$, the obtained map reflects the distribution of HI atoms at the redshift $z = \nu_{21}/{\nu_0}-1$ where $\nu_{21}$ is the frequency corresponding to wavelength 21-cm, i.e. $\nu_{21}=1.42$GHz. 
Thus the imaging tells us the three-dimensional distributions of neutral hydrogen at high redshifts. 
This is a reason why we are trying to do the imaging analysis, although it is more challenging than the power spectrum analysis.  
As for the imaging analysis, the rms noise of the brightness temperature for a bandwidth $\Delta \nu$ can be given by 
\begin{eqnarray}
	\Delta T_{\rm b} &\approx& 0.05 + 0.66\left( \frac{1+z}{8.5}\right)^{2.55}  
	\left(\frac{\Delta \nu}{1{\rm MHz}}\frac{t_{\rm int}}{1000{\rm hr}}\right)^{-1/2} \nonumber \\
	&& \times \left(\frac{\theta_{\rm b}}{7'}\right)^{-2} \rm [mK], 
\end{eqnarray}
where $\theta_{\rm b}$ is the angular resolution, and $t_{\rm int}$ is the integration time \citep{2015aska.confE..15W}. 

Here we briefly present some studies on the imaging analysis proposed in the international SKA science book. 
By semi-numerical simulations, \citet{2013MNRAS.428.3366K} have shown that the size distribution of HII bubbles is sensitive to the typical mass of galaxies providing ionizing photons to the IGM. 
Since the typical mass is determined by the efficiencies of feedback processes, the imaging analysis will bring new insights into the feedback processes during the EoR. 
Although the EoR imaging mainly focuses on large HII bubbles ($R\gtrsim20$Mpc), it will be interesting to clarify the distribution of $\delta T_{\rm b}$ around a UV source accompanied by a X-ray source. 
\citet{2008ApJ...684...18C} have simulated the emergent signatures in that case assuming a simplified density field, and shown that a mostly ionized region ($\delta T_{\rm b}\approx 0$) surrounded by a X-ray heated region ($\delta T_{\rm b}>0$) and a further Ly$\alpha$ pumping region ($\delta T_{\rm b}<0$) appear. 
\citet{2015ApJ...802....8A} have also elucidated the properties of the 21-cm signature around a rare-peak halo in a cosmological simulation and concluded that the SKA may enable us to distinguish the types of ionizing sources in such a halo. 

\subsection{21-cm forest analysis}\label{forest}
When we observe a spectrum of a high-$z$ radio loud source, foreground 21-cm absorption lines are expected to be imprinted on the spectrum. 
The 21-cm absorption feature reflects the HI distribution along the line-of-sight \citep{2002ApJ...579....1F, 2015aska.confE...6C}. 
This analysis is similar to Ly$\alpha$ forest except for using 21-cm absorption lines instead of Ly$\alpha$, thus is called 21-cm forest analysis. 
The observed flux can be written as $S_{\rm obs}=(1-e^{-\tau_{21}})S_{\rm in}$, where $\tau_{21}$ is the optical depth of the 21-cm line and $S_{\rm in}$ is the intrinsic flux of the background radio source. 
Previous studies on the 21-cm forest have shown that strong absorption features are mainly caused by small scale structures (few tens of physical kpc) that cannot be resolved by other approaches utilizing 21-cm lines \citep{2002ApJ...577...22C, 2009ApJ...704.1396X}. 
Therefore the 21-cm analysis is the unique technique which can probe the very small scale structures, but there is a big difficulty in detecting 21-cm forest. 

The minimum flux density of the background source required for the observation can be written as
\begin{eqnarray}
S_{\rm min} =\frac{2k_{\rm B}T_{\rm sys}}{A\sqrt{\Delta \nu t_{\rm int}}}\frac{S}{N}
\end{eqnarray}
where $S/N$, $k_{\rm B}$, $\Delta \nu$, $A_{\rm eff}$, and $T_{\rm sys}$ are the signal-to-noise ratio, the Boltzmann constant, the frequency resolution, the collecting area of the array, and the system noise, respectively. 
\citet{2015aska.confE...6C} have shown that $S_{\rm min}$ roughly corresponds to 10~mJy assuming $S/N=5$, $\Delta \nu = 5$~kHz, $A_{\rm eff}/T_{\rm sys}$~=~1000~$\rm m^2K^{-1}$, and $t_{\rm int}=$1000~hours for the SKA-1 low. 
Although four times better $A_{\rm eff}/T_{\rm sys}$ with the SKA2 is expected to reduce $S_{\rm min}$, there is a large uncertainty in the abundance of such luminous radio sources during or before the EoR. 
If first stars were massive as theoretically predicted and caused GRBs at their death, the GRBs might be bright enough to use for the background light of the 21-cm forest \citep{2011ApJ...731..127T}. 

\subsection{Cosmology in the EoR/CD Eras}
In the cosmological context, the SKA provides us with cosmological information of previously unobserved high redshift universe in an unprecedented spatial resolution, and thus it is a unique tool to probe many unsolved problems in cosmology.

The SKA observes the brightness temperature of the HI 21-cm line as a tracer of cosmological density field. In a very idealized situation where spin temperature is much higher than that of CMB and the gas is neutral, the brightness temperature is a unbiased tracer of the cosmological density field. 
Under such situations, SKA has a better capability to observe the density fluctuation than the Planck satellite, especially on smaller scales \citep{2008PhRvD..78f5009P,2009PhLB..673..173B}. 

In the actual observations of HI 21-cm lines, however, we have many astrophysical processes such as reionization and heating of the IGM by astrophysical objects which affect the brightness temperature in a complicated manner. In order to discriminate the astrophysical contribution from the cosmological signals, theoretical or phenomenological models for these astrophysical processes is useful \citep{2008PhRvD..78b3529M}. 
The redshift space distortion of the power spectrum of brightness temperature is also useful as a cosmological probe \citep{2006ApJ...653..815M, 2008PhRvD..78b3529M}.

One of the interesting cosmological targets of SKA is the nature of DM. 
As an alternative to CDM which has a possible deficiencies on sub-Mpc scales, warm dark matter (WDM) is a new candidate of DM with smaller mass ($\sim$ keV) and is expected to suppress structures on small scales. 
This suppression is more significant in the earlier universe, and delays the star/galaxy formation in high redshift universe. 
This feature is expected to be observed with the SKA through astrophysical contribution to the HI 21-cm signals \citep{2014MNRAS.438.2664S}. Annihilation of DM which heats the IGM uniformly, unlike the heating by astrophysical objects, can be also addressed by the HI 21-cm signals \citep{2014JCAP...11..024E}.

\section{CD/EoR Science by Japanese Researchers}\label{JAPAN}
\subsection{GYUDON Project}
\begin{figure*}
	\begin{center}
	\includegraphics[width=0.95\hsize]{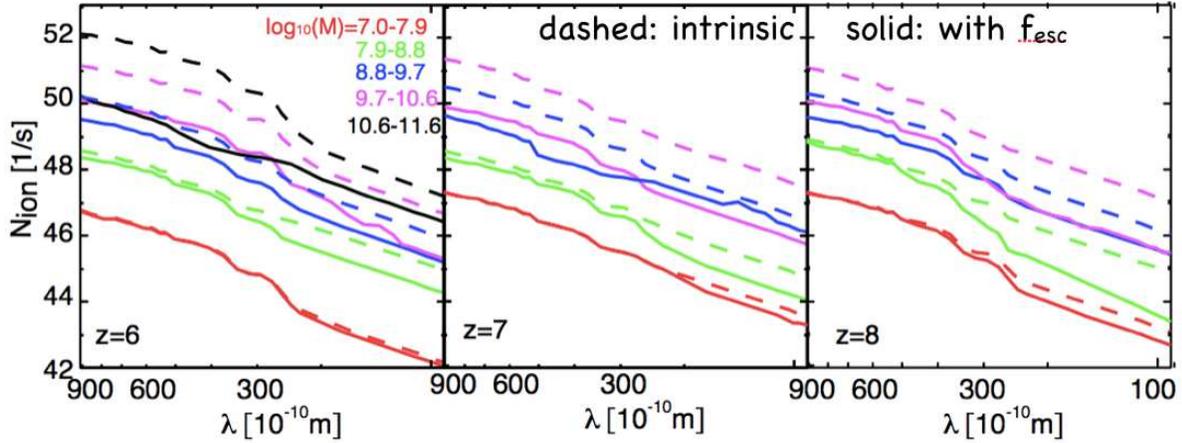} 
 	\end{center}
	\caption{UV continuum SEDs of simulated galaxies. The dashed and solid lines 
	respectively indicate the intrinsic SEDs and the emergent SEDs taking $f_{\rm esc}$ into account. 
	The red, green, blue, magenta and black lines respectively indicate the SEDs for the galaxies with 
	halo masses of $10^{7.0-7.9}M_{\odot}$, $10^{7.9{\it -}8.8}M_{\odot}$, 
	$10^{8.8-9.7}M_{\odot}$, $10^{9.7-10.6}M_{\odot}$ and $10^{10.6-11.6}M_{\odot}$. 
	Taken from Hasegawa, Ishiyama, Inoue in preparation.}
	\label{fig:SED}
\end{figure*}
As shown in \S\ref{SB}, the SKA will provide us with valuable information on the early phase of structure formation and the cosmic reionization process. 
However we need some tools for getting understandings on what we want to know from observations (see \S\ref{modelling}).  
For the purpose, we are developing a novel semi-numerical code GYUDON\footnote{Gyudon (beef bowl) is a kind of Japanese fast food which is popular for its high cost performance.}~(Galaxies and the Young Universe: DO-it-yourself Numerically), which is designed to forecast HI 21-cm signals during the CD/EoR. 
GYUDON will allow us to compute the structure formation during the EoR, spatial distributions of HII bubbles generated by the ionizing sources and the corresponding maps of 21-cm signals. 
In GYUDON, we pay special attention to the models of galaxies and IGM so that we precisely forecast the 21-cm signals. 

As described in \S\ref{physics} and \S\ref{modelling}, the feedback on the SFRs and $f_{\rm esc}$ of galaxies is crucial for calculating the reionization history, however the ionizing photon emissivities of galaxies have been frequently treated with simple assumptions. 
Thus we firstly construct a model of galaxies by analyzing the results of RHD simulations (\cite{2013MNRAS.428..154H}, Hasegawa in prep.).  
With the model, the dependences of intrinsic  ionizing photon emissivity and $f_{\rm esc}$ on the halo mass, that are regulated by the radiative and SN feedback effects, can be automatically considered.  
Fig.~\ref{fig:SED} shows the numerically obtained spectral energy distributions (SEDs) of galaxies at $z=$6, 7 and 8. 
As shown by the figure, the intrinsic UV continuum luminosity monotonically increases with the halo mass, while the escape fraction shows the opposite trend. 
Although these dependences probably affect the size distribution of HII bubbles, the dependences have often been neglected in previous simulations. 
It is worth to emphasize that the simulated UV ($1500$\AA) luminosity functions at $z>6$ are well consistent with observed ones (e.g., \cite{2015ApJ...803...34B}). 
However we should recall that there is no consensus on the absolute value of $f_{\rm esc}$ during the EoR (\S\ref{physics}). 
Thus free choice of $f_{\rm esc}(M_{\rm halo})$ as well as that derived from the RHD simulations are allowed in GYUDON. 
In GYUDON, the SED model is applied to each halo according to the halo mass in order to determine the number of ionizing photons. 

\begin{figure}
	\begin{center}
 	\includegraphics[width=0.99\hsize]{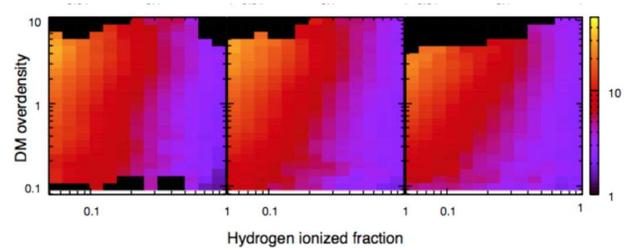} 
	\end{center}
	\caption{Look-up tables of the HII clumping factor $C_{\rm HII}\equiv \langle \rho_{\rm HII}^2 \rangle/\langle \rho_{\rm HII} \rangle^2$ at $z=9$ (left), $z=8$ (center) and $z=7$ (right). 
	 Each table is expressed as a function of local DM overdensity and hydrogen ionized fraction. 
	 The color denotes the value of $C_{\rm HII}$. 
	 Notice that there are no available data corresponding to the blank fields in the panels. 
	 Taken from Hasegawa, Ishiyama, Inoue in preparation.}
\label{fig:CHII}
\end{figure}
The distribution of matters at a given redshift is often computed by LPTs in which individual DM halos usually cannot be resolved. 
Since ionizing photons emissivities are assigned to individual halos in GYUDON, the use of $N$-body simulation results in which individual halos are resolved is preferred. 
An alternative choice for generating the matter distribution is to use COmoving Lagrangian Acceleration (COLA) method \citep{2013JCAP...06..036T}. 
In addition to enough high resolution to resolve halos, we have to model the baryonic component of the IGM since only the distribution of DM can be obtained by $N$-body simulations or COLA. 
The clumping factor defined as $\langle \rho^2 \rangle/\langle \rho \rangle^2$ of DM is generally thought to monotonically increase as the time goes by (e.g., \cite{2006MNRAS.369.1625I}), but at the same time it is almost certain that gaseous components are smoothened by photo-heating (e.g., \cite{2009MNRAS.394.1812P}). 
The smoothing leads to the modulation of effective recombination rate by changing the clumping factor. 
In order to consider the feedback on the IGM clumping factor appropriately, we make a model for it by analyzing RHD simulations (Hasegawa, Ishiyama, Inoue in prep.). 
As shown in Fig.\ref{fig:CHII}, we express the HII clumping factor $C_{\rm HII}\equiv \langle \rho_{\rm HII}^2 \rangle/\langle \rho_{\rm HII} \rangle^2 $as a function of ionized fraction and DM overdensity, based on the facts that the clumpiness is higher at dense regions and the smoothing is mainly caused by photo-heating. 
In GYUDON, the HII clumping factor table is referred every timesteps according to the local density and ionized fraction for evaluating the local recombination rate appropriately. 

\begin{figure}
	\begin{center}
	\includegraphics[width=0.99\hsize]{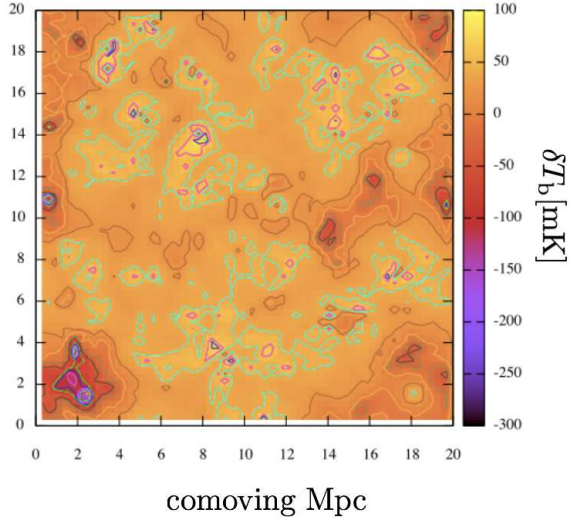} 
	\end{center}
	\caption{Contour map of $\delta T_{\rm b}$  in a comoving $(20{\rm Mpc})^3$ box at $z\approx8$ 
	computed by RT simulations utilizing 
	the models of galaxies and of IGM described in the text. 
	The color denotes $\delta T_{\rm b}$. 
	Taken from Hasegawa, 
	Ishiyama, Inoue in preparation. }
\label{fig:MAP}
\end{figure}

Fig.\ref{fig:MAP} shows the map of $\delta {T_{\rm b}}$ in a small 20 comoving Mpc cubed box, obtained  by a trial RT simulation utilizing the models of galaxies and IGM described above. 
We are planning to perform further RT/semi-numerical simulations with a larger box, several hundreds comoving Mpc on a side, for the direct forecast of 21-cm signals during the EoR. 
Here note that the progress of ionization will be solved by a semi-numerical approach instead of RT calculations in GYUDON. 

\begin{figure}
	\begin{center}
	\includegraphics[width=0.9\hsize]{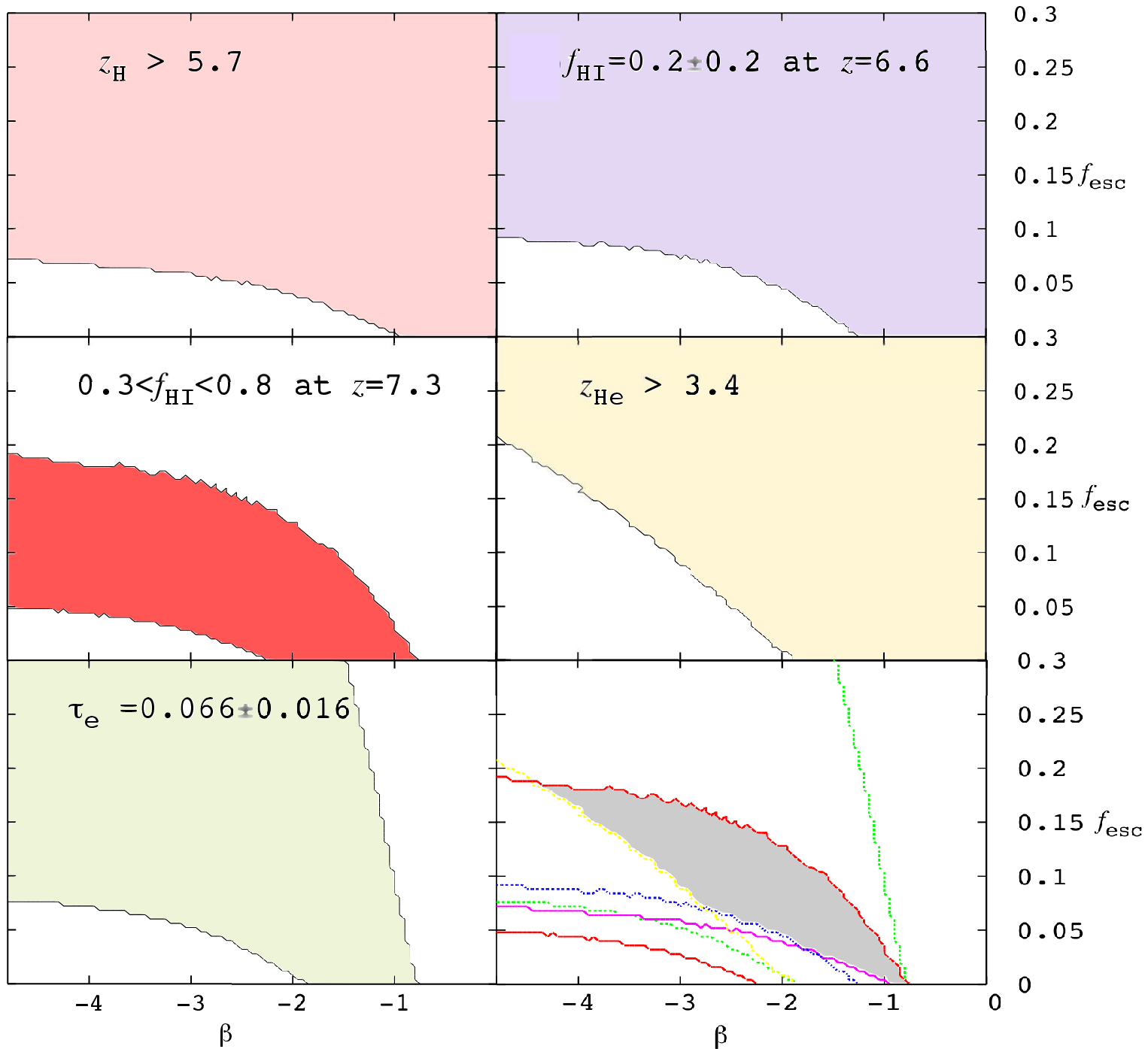} 
	\end{center}
	\caption{Observationally allowed regions in the $f_{\rm esc}$-$\beta$ parameter space. 
	In each panel, the colored region indicates the parameter space satisfying each observation. 
	From left to right,  top to bottom, each panel shows the constraint imposed by $f_{\rm HI}$ at 
	$z=5.7$ \citep{2006ARA&A..44..415F}, $f_{\rm HI}$ at $z=6.6$ \citep{2010ApJ...723..869O}, 
	$f_{\rm HI}$ at $z=7.3$ \citep{2014ApJ...797...16K}, $f_{\rm HeII}$ at $z=3.4$  
	\citep{2014arXiv1405.7405W}, and the Thomson optical depth $\tau_{\rm e}$
	\citep{2015arXiv150201589P}. 
	In the bottom-right panel, the shaded region indicates the parameter range satisfying all of
	the conditions. Taken from Fig.~6 in \citet{2016arXiv160204407Y}. ``Constraining the contribution of galaxies and active galactic nuclei to cosmic reionization", Yoshiura, S., Hasegawa, K., Ichiki, K., Tashiro, H., Shimabukuro, H., Takahashi, K., 2016, submitted to MNRAS}
 \label{fig:XRB}
 \end{figure}
In the trial calculation described in the previous paragraph, we ignored radiation from AGNs, because the number of AGNs during the EoR is still highly uncertain. 
Due to the uncertainty, it was difficult to consider the appropriate amount of UV/X-ray radiation from AGNs in previous simulations. 
In the last decade UV/X-ray radiation from AGNs was usually thought to have little contribution to the ionization process, although its impact on the signature of HI 21-cm signals due to X-ray heating has been pointed out (e.g., \cite{2010A&A...523A...4B}). 
The AGN abundance at $z\lesssim4$ is well known and is thought to be major contributor to X-ray Background (XRB) at energies below $\sim$8 keV (e.g., \cite{2014ApJ...786..104U}), whereas AGN abundance at $z\gtrsim4$ is less well-known. 
Although it was generally believed that the number of AGNs drastically decreases towards higher redshifts, \citet{2015A&A...578A..83G} have recently found a higher AGN abundance at $z\sim4-6$ by counting the number of faint AGNs. 
If such a high AGN abundance is valid beyond $z\sim6$, AGNs might have significant contribution to reionization. 

In \citet{2016arXiv160204407Y}, we succeeded in constraining AGN abundance at $z\gtrsim3$ by comparing numerically computed  HI/HeII reionization histories to several observations regarding HI/HeII fractions at various redshifts.  
Assuming that the AGN abundance evolves as $(1+z)^{\beta}$ at $z>3$, we find that $\beta$ should lie between $\sim-4$ and $\sim-3$ with the ionizing photon escape fraction of $f_{\rm esc} \sim0.2$ (Fig. \ref{fig:XRB}). 
Interestingly, if the high AGN abundance claimed by \citet{2015A&A...578A..83G} is valid at $z>6 $ and the escape fraction of ionizing photons from high-$z$ galaxies is as small as 1\%, the numerical model can satisfy all the observational constraints regarding HI/HeII fractions. 
It simultaneously means that there is a degeneracy between $f_{\rm esc}$ and $\beta$. 
Unfortunately it is difficult to resolve the degeneracy by ongoing observational instruments, but the SKA will be able to resolve it by bringing the map of $\delta T_{\rm b}$. 
The constrained AGN model made in this work will be implemented to GYUDON so as to include the effects of AGNs as far as the current observational constraints allow. 

\subsection{Bispectrum and its detectability}
\begin{figure}
  \begin{center}
   \includegraphics[width=0.9\hsize]{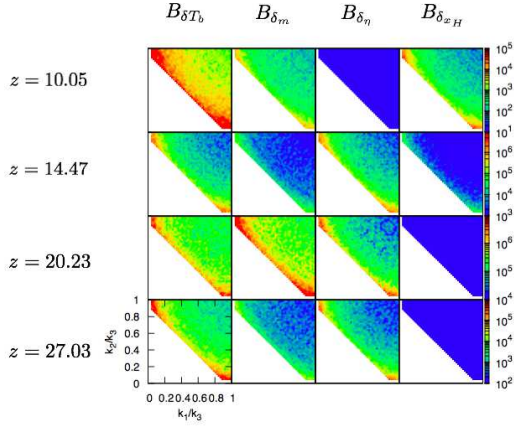} 
  \end{center}
 \caption{Contours of the total bispectrum and its component in $k_{1}/k_{3}-k_{2}/k_{3}$ plane fixing $k_{3}=1.0{\rm Mpc}^{-1}$. Here, we plot the unnormalized bispectrum in units of $\rm Mpc^{6} mK^{3}$. 
 From left to right, each one describes the bispectrum for brightness temperature, matter fluctuation, spin temperature and neutral hydrogen at each redshift. Taken from Fig.~5 in \citet{2015arXiv150701335S}. ``21cm-line bispectrum as method to probe Cosmic Dawn and Epoch of Reionization", Shimabukuro, H., Yoshiura, S., Takahashi, K., Yokoyama, S., Ichiki, K., 2016, accepted in MNRAS}
 \label{fig:bs_contour}
 \end{figure}
Since mapping the brightness temperature requires high spatial resolution, it is important to study the properties of the 21-cm fluctuations statistically as a first step. 
In \citet{2015MNRAS.451..467S}, we studied the properties of the variance and skewness of the 21-cm fluctuations and found that the skewness is a good indicator of the epoch when X-ray heating starts working effectively. 

However, the skewness does not tell us the information on the scale dependence of 21-cm fluctuations, since the skewness is obtained by integrating bispectrum over wavenumbers.
Thus we need to study the bispectrum of 21-cm fluctuations to assess the scale dependence. 
Furthermore, the bispectrum analysis is useful to investigate non-Gaussian features that are likely caused by the various astrophysical effects such as the WF effect, X-ray heating and reionization. 
Despite the importance of the bispectrum analysis, the properties of the bispectrum of the 21-cm fluctuations have never been elucidated. 
Therefore in \cite{2015arXiv150701335S}, we firstly focus on the bispectrum of the 21-cm signals.  
The 21-cm bispectrum is defined by 
\begin{eqnarray}
\langle\delta_{21}(\boldsymbol{ k_{1}}) \delta_{21}(\boldsymbol{ k_{2}}) \delta_{21}(\boldsymbol{ k_{3}})\rangle &=& (2 \pi)^3 \delta(\boldsymbol{ k_{1}}+\boldsymbol{ k_{2}}+\boldsymbol{ k_{3}}) \nonumber \\
&& \times B(\boldsymbol{ k_{1}},\boldsymbol{ k_{2}},\boldsymbol{ k_{3}}),
\label{eq:bs_def}
\end{eqnarray}
where, $\delta_{21}=\delta T_{b}-\overline{\delta T_{b}}$, and $\overline{\delta T_{b}}$ is the mean brightness temperature.
We compute maps of the 21-cm brightness temperature by utilizing 21CMFAST \citep{2011MNRAS.411..955M}, and calculate the 21-cm bispectrum with respect to the maps.
As a result, we find that the 21-cm bispectrum is almost scale invariant at $z\approx$ 10, 14 and 27. 
On the other hand, it increases with wavenumber at $z\approx20$. 
This is because the dominant component of 21-cm bispectrum is different at each epoch.
In order to see what the dominant component in  the 21-cm bispectrum is, we decompose the 21-cm bispectrum. 
Here we define a new parameter $\eta=1-T_{\rm CMB}/T_{\rm s}$ and rewrite Eq.~(\ref{dts}) as 
\begin{equation}
	\delta T_{\rm b} = \overline{\delta T_{\rm b}}(1+\delta_{m})(1+\delta_{x_{\rm H}})(1+\delta_{\eta}), 
\label{ddd}	
\end{equation}
where $\delta_{m}$, $\delta_{x_{\rm H}}$ and $\delta_{\eta}$ are respectively the fluctuations of the matter, neutral hydrogen fraction and $\eta$. 
By using Eq.~(\ref{ddd}) we decompose the 21-cm bispectrum into the auto- and cross-correlations of $\delta_{m}$, $\delta_{x_{\rm H}}$ and $\delta_{\eta}$; 
\begin{eqnarray}
B_{\delta T_b}
&=& (\overline{\delta T}_{b})^{3}
    [B_{\delta_m }
     + B_{\delta_{x_{\rm H}}}
     + B_{\delta_{\eta}} \nonumber \\
& &  + ({\rm cross ~ correlation ~ terms}) \nonumber \\
& &  + ({\rm higher ~ order ~ terms})], 
\label{eq:bispectrum_component}
\end{eqnarray}
where $B_{\delta_m}$, $B_{\delta_{x_{\rm H}}}$ and $B_{\delta_\eta}$ denote the auto-correlation terms of $\delta_{m}$, $\delta_{x_{\rm H}}$ and $\delta_{\eta}$, respectively. 
Fig.~\ref{fig:bs_contour} shows each component of the 21-cm bispectrum at $z\approx$10, 14, 20, and 27. 
As shown by Fig.~\ref{fig:bs_contour}, the bulk of the 21-cm bispectrum at $z\approx$20 comes from both the matter and spin temperature fluctuations, while at the other epochs those are mainly determined by the spin temperature and/or the neutral fraction.  
It is expected that we obtain more detailed information on the IGM during the CD/EoR by using the 21-cm bispectrum in the future experiments, combined with the power spectrum and skewness.

  \begin{figure*}
  \begin{center}
   \includegraphics[width=0.8\hsize]{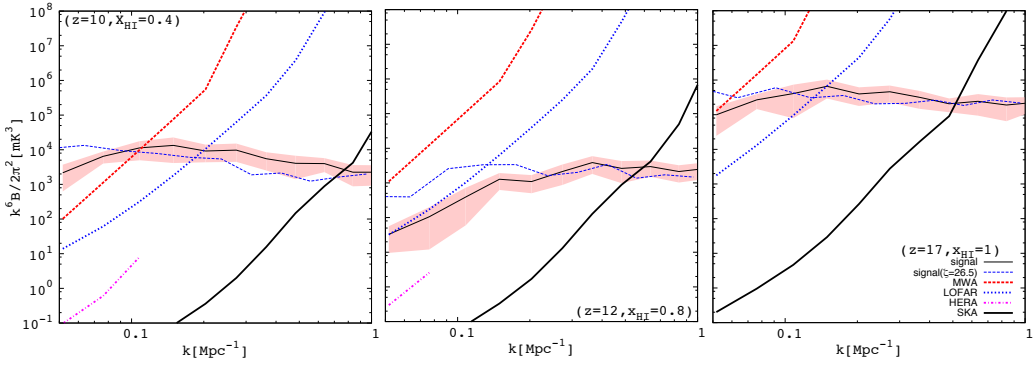} 
  \end{center}
 \caption{Comparison of equilateral-type bispectrum signal of variant models with ionizing efficiency $\zeta = 31.5$ and $\zeta = 26.5$ (thin solid line and thin dashed line), sample variance of signal (shaded area) and thermal noise  of MWA (thick dashed), LOFAR (thick dotted), HERA(thick dot dashed) and SKA (thick solid) at $z = 10, 12$ and $17$ from left to right. Here $x_{\rm HI}$ is the neutral fraction of the fiducial model at each redshift. Taken from Fig.~1 in \citet{2015MNRAS.451..266Y}. ``Sensitivity for 21 cm bispectrum from Epoch of Reionization", Yoshiura, S., Shimabukuro, H., Takahashi, K., et al. 2015, MNRAS, 451, 266}
 \label{fig:Y_noise}
 \end{figure*}

As described above, the bispectrum is a promising observable to probe the CD/EoR, but its detectability was not investigated. 
Therefore in \citet{2015MNRAS.451..266Y}, we estimated the thermal noise for the bispectrum observation. 
Because the thermal noise follows a Gaussian distribution, the ensemble average of the noise bispectrum vanishes and the variance of the noise bispectrum contributes to the bispectrum signal. We developed a formalism to calculate the noise bispectrum for an arbitrary triangle, taking the array configuration into account. 
Defining $K \equiv |{\boldsymbol{k}}_1|$ and $k \equiv |{\boldsymbol{k}}_2| = |{\boldsymbol{k}}_3|$, the bispectrum variance due to the thermal noise is estimated as 
\begin{eqnarray}
& &\delta B_N(k,K)
= \frac{(2\pi)^{\frac{5}{2}}V}{\sqrt{\Delta \theta_2} k K^{3/2} \epsilon}
    \bigg(\frac{T^2_{\rm sys} \lambda^2}{A_{\rm eff} B t_{\rm int}}\bigg)^{\frac{3}{2}} 
\nonumber \\
& & \times \left[ \int {d\theta_1} \int {d\alpha} \sin{\theta_1} \sin{\theta_2} 
                 \sin{\gamma} 
                 ~ n({\boldsymbol{k}}_1) n({\boldsymbol{k}}_2) n({\boldsymbol{k}}_3) \right]^{-\frac{1}{2}},
\end{eqnarray}
where $\theta_i$ is the polar angle of ${\boldsymbol{k}}_i$, $\alpha$ is the rotational degree of freedom for ${\boldsymbol{k}}_2$ with respect to ${\boldsymbol{k}}_1$, $\gamma$ is the angle between $\partial {\boldsymbol{k}}_2/\partial \alpha$ and $\partial {\boldsymbol{k}}_2/\partial \theta_2$, $\lambda$ is the observed wavelength, $V$ is the observed volume in real space, $B$ is the bandwidth, $\epsilon$ is the constant factor set equal to 0.5, and $n({\boldsymbol{k}})$ is the number density of baselines. For equilateral type, we just set $K = k$.
We applied it to the cases with equilateral type of bispectrum and estimated the noise bispectrum for Murchison Widefield Array (MWA), LOw Frequency ARray (LOFAR), Hydrogen Epoch of Reionization Array (HERA) and the SKA. Consequently, as shown in Fig.\ref{fig:Y_noise}, it was found that the SKA has enough sensitivity for $k \lesssim 0.3~{\rm Mpc}^{-1}$ (comoving). HERA also has enough sensitivity, however, because of short baselines, it will not observe signal on small scales. On the other hand, LOFAR and MWA have sensitivity for the peaks of the bispectrum as a function of redshift. 

\subsection{21-cm signatures around the first stars and their SN remnants}
The high sensitivity and angular resolution of SKA will allow us to investigate the 21-cm signal
around first stars. 
In \citet{2014MNRAS.445.3674Y}, we have calculated the 21-cm structure by using radiative transfer simulations that include ionization of hydrogen and resonant scattering of Ly$\alpha$ photons.
A striking feature of this work compared to other studies shown in \S\ref{imaging} is to solve Ly$\alpha$ radiative transfer that estimates number of scatterings of Ly$\alpha$ photons for considering the WF effect precisely, leading to more appropriate evaluation of the 21-cm spin temperature. 
Our models are based on one-dimensional spherical shells with the uniform IGM density $n_{\rm H}^{\rm IGM} \sim 2 \times 10^{-7} (1+z)^{3}~\rm cm^{-3}$.
We follow the time-evolving structures of ionization, temperature and radiation field of the central first stars, and then estimate the spin temperature and the 21-cm signal. 
Although the life time of first stars is $\sim$ several Myr, our calculations follow the time evolution until $10^8$ yr considering continuous star formation of 
multiple stars.   

\begin{figure*}
  \begin{center}
   \includegraphics[width=0.8\hsize]{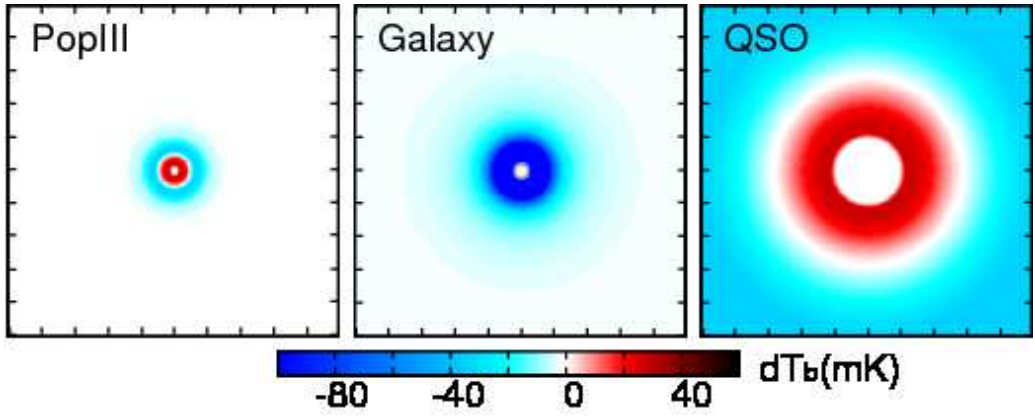} 
  \end{center}
 \caption{Two-dimensional map of the differential brightness temperature of first stars ($M_{\rm star}=200~\rm M_{\rm \odot}$), galaxy ($M_{\rm star}=2\times10^{6}~\rm M_{\rm \odot}$) and QSO ($M_{\rm star}=2 \times 10^{8}~\rm M_{\rm \odot}$) at the evolution time of $10^{7}~\rm yr$. The box size is 500 kpc in physical scale. Taken from Fig.~5 in \citet{2014MNRAS.445.3674Y}. ``Distinctive 21-cm structures of the first stars, galaxies and quasars", Yajima, H., Li, Y. 2014, MNRAS, 445, 3674}
 \label{fig:map_Yajima14}
 \end{figure*}

The left panel of Fig.~\ref{fig:map_Yajima14} shows the simulated 21-cm signal around first stars with $M_{\rm star}=200~M_{\rm \odot}$ at $z=10$.
The differential brightness temperature shows the ring-like structure transiting from a positive value to a negative one. 
The central hole represents almost completely ionized regions. 
The high-energy photons from first stars produce the large transition area from ionized to neutral regions
 where the gas temperature is higher than the CMB temperature. 
Due to strong Ly$\alpha$ radiation from halos,
the spin temperature in the transition area is coupled to the gas temperature. 
As a result, the transition area shows an emission signal. 
The absorption signal, which represents the cold neutral region, becomes weaker as the radial distance increases.
We also calculate the 21-cm signal around galaxies and quasars of which the shape of SED and the stellar mass are different. 
The feature of 21-cm signal of QSOs (right panel) is similar to that of the first stars, but the size of the 21-cm signal is much bigger due to higher luminosity. 
On the other hand, the 21-cm signal of galaxies (central panel) shows the small emission signal region due to lower high-energy photon flux ratio.
We find that the peak positions of the emission and absorption signals around the first stars ($M_{\rm star}=200~M_{\rm \odot}$) are $r=10.0$ kpc ($\delta T_{\rm b} = 28.0$ mK) and $r=50.1$ kpc ($\delta T_{\rm b} = -2.5$ mK) at $t=10^{6}$ yr,
then move to $r=17.8$ kpc ($\delta T_{\rm b} = 27.7$ mK) and $r=85.8$ kpc ($\delta T_{\rm b} = -70.1$ mK) at $t=10^{7}$ yr. 

\begin{figure}
  \begin{center}
   \includegraphics[width=0.9\hsize]{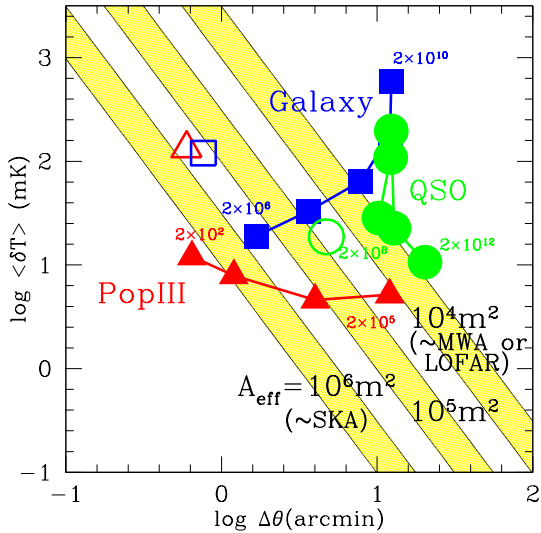} 
 \end{center}
 \caption{Detectability of the 21-cm signal from the first stars, the galaxies, and the first quasars with upcoming missions MWA, LOFAR, and SKA. 
 The filled symbols represent $\delta T_{\rm b}$ at $z=10$ with different stellar masses
 which is estimated in the spherical top-hat beam with the size of the distance from the source to the outer edge of $|\delta T_{\rm b}|=1~\rm mK$. 
 The numbers near the symbols show the stellar masses in the unit of solar mass. 
 The open symbols are the mean $\delta T_{\rm b}$ of the lowest stellar masses for each source at $z=20$.
 The yellow shaded regions indicate the sensitivity for $z=10$ by the forthcoming facilities with the effective collect areas of $10^{4}, 10^{5}$
 and $10^{6}~\rm m^{2}$, 
 while its width corresponds to the integration time from 100 to 1000 hours with $\Delta \nu = 1~\rm MHz$. Taken from Fig.~10 in \citet{2014MNRAS.445.3674Y}. ``Distinctive 21-cm structures of the first stars, galaxies and quasars", Yajima, H., Li, Y. 2014, MNRAS, 445, 3674}
 \label{fig:observability}
 \end{figure}

The detailed structure of the 21-cm signal cannot be resolved even by the SKA. 
Hence, we estimate  $\langle \delta T \rangle$ by making the coarse-grained images of the 21-cm signals with the radius where $|\delta T_{\rm b}|$ becomes less than 1 mK. 
If the UV background is not strong and the separation distance of halos is much larger than the angular resolution, 
SKA will be able to resolve the rough position of first stars. 
The detectability of the 21-cm signal around first stars, galaxies and QSOs are shown in Fig.~\ref{fig:observability}.
Different symbols represent the results of the different stellar masses.
The yellow shade regions show the sensitivity curves of the telescopes with the effective correct areas of $10^{4},~10^{5}$ and $10^{6}~\rm m^{2}$, and
their lower and upper lines represent the integration time of 1000 and 10000 hours, respectively.
The size $\Delta \theta$ increases with the stellar mass and the size of H{\sc ii} bubbles. 
At $z=10$, Pop III stars show $\langle \delta T \rangle \sim 5 - 12~\rm mK$, while galaxies have $\langle \delta T \rangle \sim 19 - 584~\rm mK$ and QSOs have $\langle \delta T \rangle \sim 28 - 196~\rm mK$. 
We suggest that the 21-cm signal around the cluster of first stars with $M_{\rm star} \sim 10^{4}~\rm M_{\rm \odot}$ at $z=10$ can be detected by SKA1-low with an integration time of $\sim 1000$ hours.

First stars cause supernovae~(SNe) at their death because first stars can be enough massive. 
Therefore, SNe of first stars are expected as observational signature of first stars. 
In particular, first stars in the mass range $140~M_\odot< M < 260~M_\odot$ explode as pair-instability SNe, which are up to 100 times more energetic than typical Type Ia and Type II SNe~\citep{2010ApJ...724..341H}.
Many authors have studies the detectability of such energetic SNe at the EoR ~\citep{2010AIPC.1294..268M,2012MNRAS.422.2675T,2013ApJ...762L...6W,2014ApJ...792...44C,2015ApJ...805...44S}.
and concluded that these SNe could be detected by future high redshift surveys including JWST~\citep{2006SSRv..123..485G} and EUCLID~\citep{2011arXiv1110.3193L}.

\begin{figure}
 \begin{minipage}{0.5\hsize}
  \begin{center}
   \includegraphics[width=70mm]{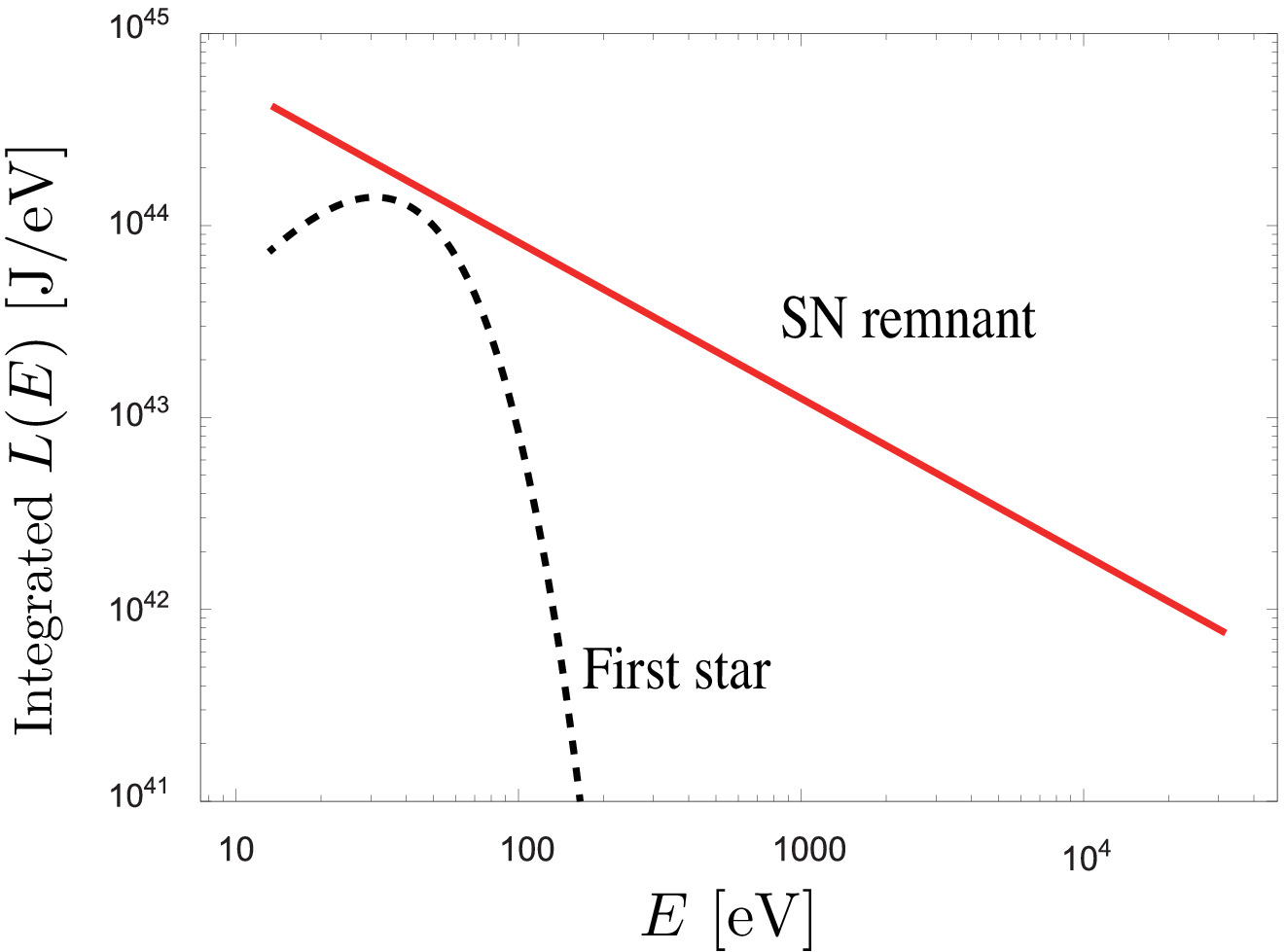}
  \end{center}
 \end{minipage}
 \begin{minipage}{0.5\hsize}
  \begin{center}
   \includegraphics[width=70mm]{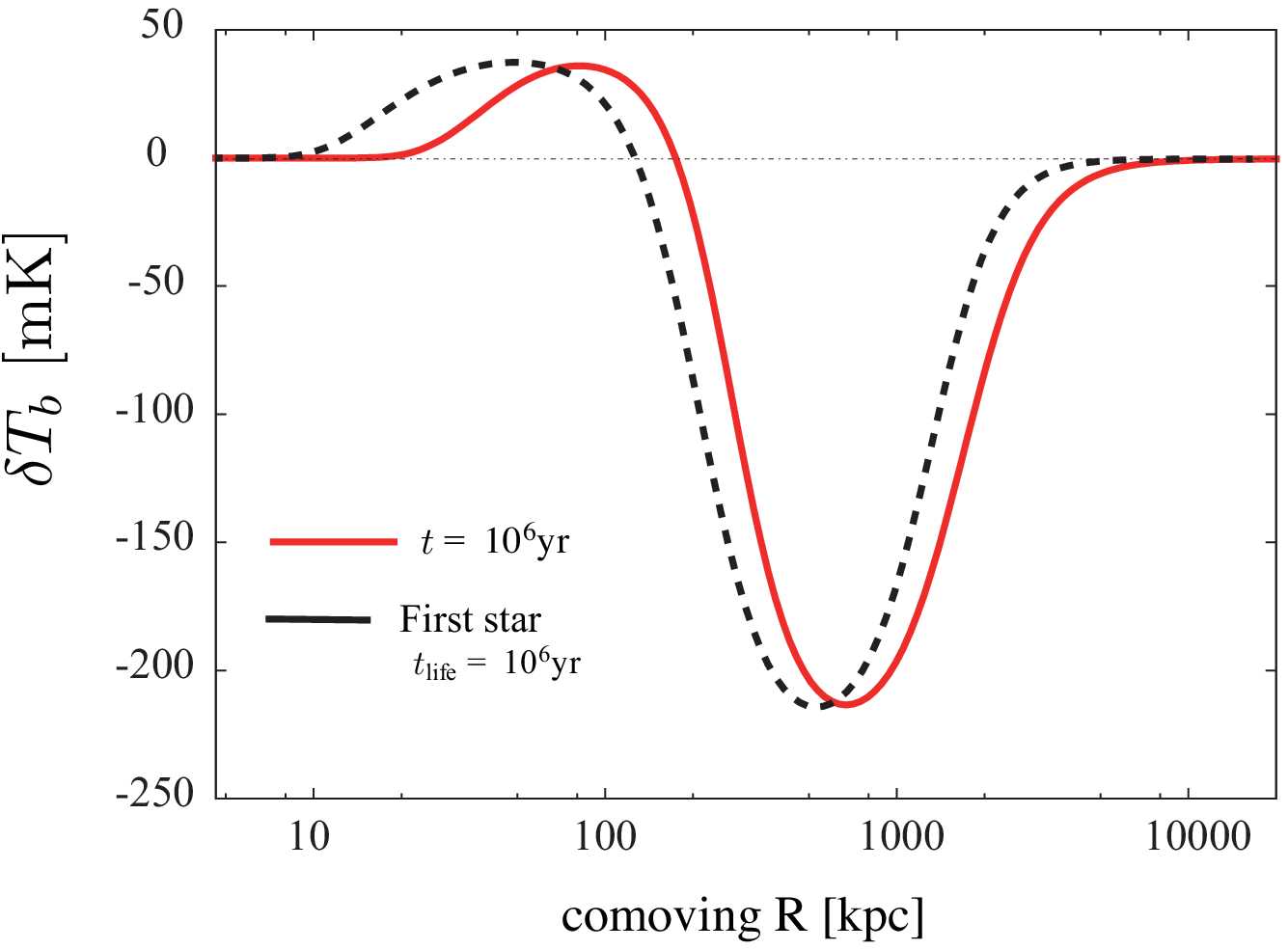}
  \end{center}
 \end{minipage}
  \caption{
 {\it Top}:~The luminosity integrated over the lifetime
 of a SN remnant~(red solid line).
 {\it Bottom}:~ The 21-cm signals due to the first star
 SN in the right~(red solid line). In both panels, we set $E_{\rm SN} = 10^{53}~\rm erg$, $f_{\rm rel} =0.01$ and
$f_{\rm mag}=0.01$. For comparison, we plot the results for the first star model as black dotted lines. Taken from Tashiro in preparation. }\label{fig:sn-21}
\end{figure}

SNe of first stars can imprint observable signatures on redshifted 21-cm maps. 
SN remnants produce X-ray photons which efficiently heat up their surroundings.
As a result, the spin temperature of IGM near SNe of first stars decouples from the CMB temperature and the observable signals due to SNe in redshifted 21-cm maps are created.
We have investigated the 21~cm signal around a SN of a first star, using a simple analytical model of a SN remnant.
As the evolution of the SN remnant, we adopt the Sedov-Taylor solution with SN energy~$E_{\rm SN}$.
The Sedov-Taylor solution is a solution in adiabatic phase. 
However, the remnant is cooled through the Compton scattering with CMB photons.
The time scale of the Compton cooling is given by $t_{\rm C} =3 m_e c /4 \sigma_T a T_{\rm CMB}^4 \sim 1.4 \times 10^7 [(1+z)/20]^{-4}~\rm yr$. 
The energy of the remnant is totally lost by the Compton cooling when $t \sim t_{\rm C}$.
In the propagation of a SN remnant through the ambient medium, electron acceleration to relativistic velocities and magnetic field amplification occur. 
In order to evaluate the relativistic electron number density and magnetic field amplitude in a SN remnant, we introduce the fractions of the SN energy into relativistic
electrons~($f_{\rm rel}$) and magnetic fields~($f_{\rm mag}$).
In the presence of magnetic fields and relativistic electrons, X-ray photons are produced through the synchrotron radiation. 
We calculate the luminosity through the synchrotron radiation from the SN remnant of a first stars with $E_{\rm SN} = 10^{53}~\rm erg$, $f_{\rm rel} =0.01$ and $f_{\rm mag}=0.01$. 
We show the luminosity integrated over the lifetime of a SN remnant $L(E)$ in the top panel of Fig.~\ref{fig:sn-21}. 

We also investigate the ionization and the heating of surrounding IGM by the X-ray photons from a SN remnant. 
As described in \S\ref{physics}, X-ray photons can ionize the IGM at a distance of several kpc, and heat the IGM up to Mpc. 
We show the resultant 21-cm signals due to the first star SN in the bottom panel of Fig.~\ref{fig:sn-21}.
The size of the detectable signal for the redshifted 21-cm lines also reaches a several comoving
Mpc scales. 
For example, the SKA noise level is roughly $50~\rm mK$ at $z\sim 20$ with 1000~hours integration. The radius of the sphere with $|\delta T| \sim  50~ \rm mK$ is $3~$comoving Mpc, whose angular scale is $\sim 3'$. 
Therefore, SKA might be able to resolve the 21-cm signals due to the SN of a first star.

\subsection{21-cm forest}
\begin{figure}
 \begin{center}
   \includegraphics[width=0.9\hsize]{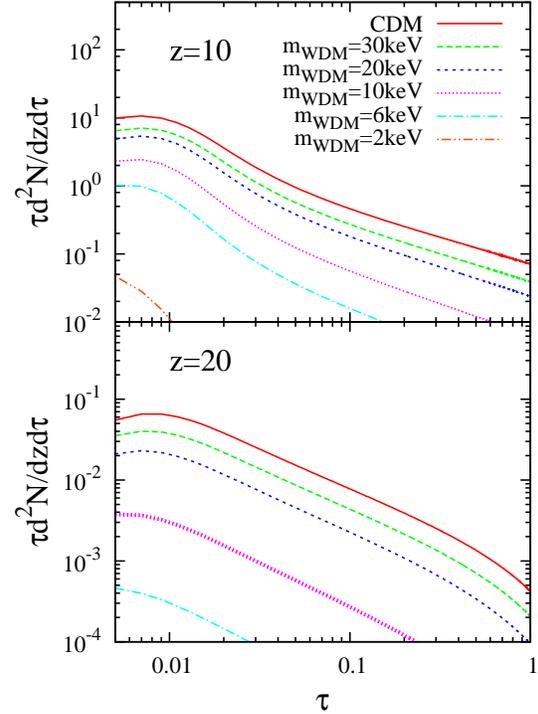} 
 \end{center}
\caption{Expected number of 21-cm absorber (N) along a line of sight, as a function of optical depth against 21-cm line ($\tau$), per redshift interval at z=10 (top) and z=20 (bottom) in the WDM cosmology for several DM masses as indicated. WDM with smaller mass erases density fluctuations on larger length scales, leading to less abundance of the absorber. Taken from Fig.~10 in \citet{2014PhRvD..90h3003S}. ``Probing small-scale cosmological fluctuations with the 21 cm forest: Effects
of neutrino mass, running spectral index, and warm dark matter", Shimabukuro, H., Ichiki, K., Inoue, S., Yokoyama, S. 2014, Phys. Rev. D, 90, 083003}
\label{fig:21cmWDM}
\end{figure}
As described in \S \ref{forest}, the 21-cm forest is an radio analogue to the Ly$\alpha$ forest, that is, systems of narrow absorption lines due to intervening, cold neutral intergalactic medium or collapsed structures against high-$z$ radio loud sources. 
An advantage of the 21-cm forest compared to the Ly$\alpha$ forest is its smaller optical depth that enables us to probe higher redshifts $z>6$. 
The 21-cm absorption lines are mainly created by small scale structures such as filaments and mini-halos in the IGM, and the number of the lines strongly depends on the thermal state of the IGM and the
number of mini-halos. 
Therefore, the 21-cm forest can be a powerful probe to the EoR and the nature of density fluctuations on the corresponding scales of mini-halos. 

Using the ability of the 21-cm forest to see through a high-redshift universe, \citet{2014PhRvD..90h3003S} show that the 21-cm forest can be used to probe the power spectrum of density fluctuations on small sales $k\gtrsim 10$ Mpc$^{-1}$ (comoving) that other cosmological probes can not reach. 
Using an analytic model by \citet{2002ApJ...579....1F}, we predict the number of absorption lines as a function of optical depth in cosmologies somewhat beyond the simplest $\Lambda$CDM model. 
As a specific example, we consider the effects of finite neutrino mass, running spectral index and WDM on the power spectrum. 
In particular, it is shown that the 21-cm forest is sensitive at the WDM mass range of $m_{\rm WDM}\gtrsim 10$ keV that may otherwise be difficult to access, for instance, through clustering analyses of galaxy distribution and the Ly$\alpha$ forest \citep{2013PhRvD..88d3502V}.
Fig.~\ref{fig:21cmWDM} shows expected number of absorption lines per redshift interval in models with different WDM masses. 
The free streaming effect of lighter WDM erases density fluctuations on larger scales leading to less mini-halos, and therefore the number of absorption lines decreases with decreasing WDM mass. 
Because the 21-cm forest is created dominantly by mini-halos on smaller scales than the Ly$\alpha$ forest can probe, it is sensitive to lighter WDM masses.

Whereas the 21-cm forest observation would offer a powerful probe to the EoR and cosmology, the most challenging issue of the detection of a 21-cm forest is the existence of radio loud sources at high redshifts. Since we can expect that the number of the absorber reaches a maximum around $\tau \approx 0.01$ as shown in Fig.~\ref{fig:21cmWDM}, we need to detect absorption lines with $\tau \gtrsim 0.01$ to use the 21-cm forest for a cosmological purpose,  
The number of the QSO sources whose flux is larger than the detection threshold of an absorption line with $\tau\approx 0.01$ is expected as $\lesssim 100$ in the whole sky at $z=10$, assuming a integration time of $10$~hours and the SKA2 sensitivity \citep{2015aska.confE...6C}. 
Another candidate for the background radio source is the radio afterglows of GRBs arising from Pop III stars \citep{2015MNRAS.453..101C}. 
In any case, if a sufficiently bright radio source was found at a high redshift the absorption lines should be easily detected with the SKA, because nagging foregrounds are not a concern for observing the 21-cm forest.

\subsection{The effect of supersonic streaming motion on structure formation}
 \begin{figure}
  \begin{center}
   \includegraphics[width=0.9\hsize]{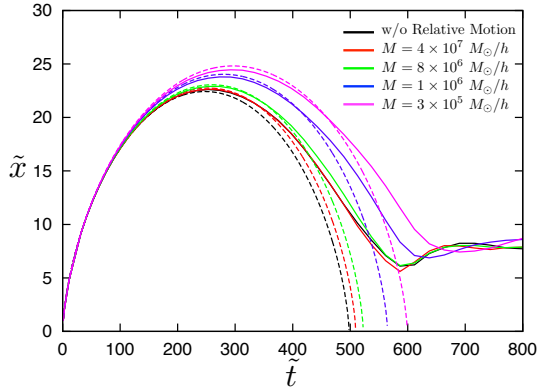} 
  \end{center}
 \caption{The time evolution of the sizes of DM halos $M=4\times10^7~M_\odot$ (red), $M_c=8\times10^6~M_\odot$ (green),
$M_c=1\times10^6~M_\odot$ (blue), and $M_c=3\times10^5~M_\odot$ (magenta) with relative velocity between DM and baryons $v_{bc}=30~{\rm km/s}$.
The black lines show the case without relative motion.
The solid lines represent the results by $N$-body simulations and the dashed lines are the numerical evaluations in the semi-analytical model. Taken form Fig.~5 in \cite{PhysRevD.93.023518}. ``Effect of supersonic relative motion between baryons and dark matter on collapsed objects", Asaba, S., Ichiki, K., Tashiro, H. 2016, Phys. Rev. D, 93, 023518}
  \label{fig_SA1}
 \end{figure}
 \begin{figure}
  \begin{center}
    \includegraphics[width=0.9\hsize]{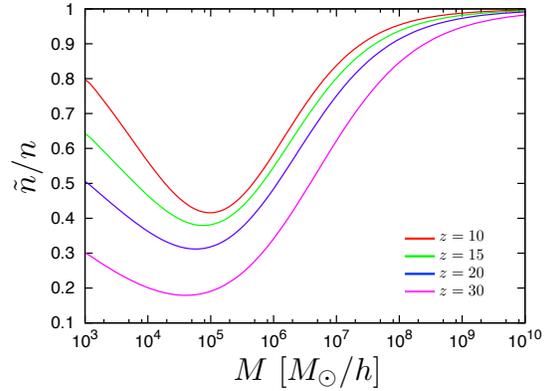} 
  \end{center}
 \caption{The ratio of DM halo abundance between with and without considering the delay of the halo formation by the supersonic streaming motions at $z=10$ (red), $z=15$ (green), $z=20$ (blue) and $z=30$ (magenta) as a function of the DM halo mass. Taken form Fig.~11 in \cite{PhysRevD.93.023518}. ``Effect of supersonic relative motion between baryons and dark matter on collapsed objects", Asaba, S., Ichiki, K., Tashiro, H. 2016, Phys. Rev. D, 93, 023518}
  \label{fig_SA2}
 \end{figure}
As described in \S\ref{physics}, the supersonic streaming motion between baryons and CDM suppresses a halo abundance on small scales, and impacts the reionization history via modifications of star formation processes.
However, the detailed process of the suppression has not been understood sufficiently. 

In \citet{PhysRevD.93.023518}, we investigated the effect of the streaming motion on the formation of DM halos in the context of the spherical collapse model.
We performed $N$-body simulations of the collapse of an isolated DM halo in a uniform background density, taking the streaming motion into account by giving the coherent relative velocity to one-sixth ($\approx \Omega_{\rm b}/\Omega_{\rm m}$) of the background particles. 
Fig.~\ref{fig_SA1} shows the time evolution of the sizes of DM halos with relative velocity $v_{bc}=30~{\rm km/s}$. 
For comparison, we also plot the solutions of a semi-analytical model given by
\begin{eqnarray}
	\frac{d^2 x_c}{dt^2}&=&-\frac{G(M_{c,i}+M_b)}{x_c^2},\nonumber\\
	M_{c,i}&=&\frac{4\pi}{3} \bar{\rho}_{c,i}x_{c,i}^3(1+\delta_{c,i}),\nonumber\\
	M_b&=&\frac{4\pi}{3}\bar{\rho}_bx_c^3(1+\delta_b). \label{rveq}
\end{eqnarray}
When we solve the above equations, $M_b$ is estimated from a $N$-body simulation result so as to consider the effect of the streaming motion. 
Fig.~\ref{fig_SA1} tells us that the collapse time of DM halos is delayed by the streaming motion and is more remarkable on smaller mass scales. 
Moreover, the comparison with the semi-analytic solution illustrates that the delay due to the streaming motion is related to the decrease in the baryon mass. 
The degree of the delay depends on not only the halo mass and the amplitude of the relative velocity but the collapse epoch of the halo in the absence of the streaming motion. 

It is convenient if we have a simple manner in which the influence of the streaming motions is taken into account. 
We have firstly modified the Press-Schechter formalism to consider the abundance of DM halos affected by the streaming motions. 
In the modified formalism, the delay of the collapse time is considered by changing the value of the critical density $\tilde{\delta}_{\rm crit}$, and the mass function is given by 
\begin{eqnarray}
	\tilde{n}(M,z)&=&\int f(v_{bc})\sqrt{\frac{2}{\pi}}\frac{\bar{\rho}(z)}{M}\frac{d}{dM}\left[\frac{\tilde{\delta}_{\rm crit}(M,v_{bc}, z)}{\sigma(M, z)}\right] \nonumber \\
	&& \times \exp\left[-\frac{\tilde{\delta}_{\rm crit}^2(M,v_{bc}, z)}{2\sigma^2(M, z)}\right]dv_{bc},
	\label{ps}
\end{eqnarray}
where $f(v_{bc})$ is the probability function of the amplitude of the relative velocity between baryons and DM at cosmic recombination.
Fig.~\ref{fig_SA2} shows the ratio of halo abundances between with and without the supersonic streaming motions.
We find the halo abundance with $M\le 10^8~M_\odot/h$ is suppressed by the streaming motions and, in particular, the halo abundance on mass scale $M\sim 10^5~M_\odot/h$ decreases  by 80\% ($z=30$) and a half ($z=10)$.
It is expected that this suppression of the halo abundance alters the reionizaiton history and affects the 21-cm forest signal produced by minihalos.
Therefore we should involve the effect of the streaming motion to simulate the 21-cm signals accurately.

\section{Summary}
The SKA will be a powerful instrument for exploring the CD/EoR Science in the nest decade. 
As described in this article, observational quantities expected to be detected by the SKA involves fruitful information on the properties of astronomical objects causing reionization as well as the cosmic reionization process itself . 
The studies by SKA-Japan Consortium members shown in \S\ref{JAPAN} have already contributed to the CD/EoR Science with the SKA, but further international collaborations will be valuable in modeling of reionization and in understanding the physical processes in the EoR. 

\begin{ack}
The authors are grateful to International SKA CD/EoR SWG members for providing us opportunities of open discussion and cooperation. 
We also thank Hiroyuki Hirashita and Daisuke Yamauchi for carefully reading our manuscript. 
This work was supported in part by Grant-in-Aid from the Ministry of Education, Culture, Sports, Science and Technology (MEXT) of Japan, No. 26610048, No. 15H05896 (K.T.), No. 24340048 (K.T., K.I.), No. 25-3015 (H.S.), No. 15K17646 (H.T.), and No. 26-2667(S.A.). 
H.T. also acknowledges the support by MEXT's Program for Leading Graduate Schools PhD professional, ``Gateway to Success in Frontier Asia''.
The work of K.H. was also supported by a grant from NAOJ. 

\end{ack}


\bibliographystyle{aa}
\bibliography{paper_v4}

\begin{thebibliography}{107}
\expandafter\ifx\csname natexlab\endcsname\relax\def\natexlab#1{#1}\fi

\bibitem[{{Abel} {et~al.}(2002){Abel}, {Bryan}, \&
  {Norman}}]{2002Sci...295...93A}
{Abel}, T., {Bryan}, G.~L., \& {Norman}, M.~L. 2002, Science, 295, 93

\bibitem[{{Adshead} \& {Furlanetto}(2008)}]{2008MNRAS.384..291A}
{Adshead}, P.~J. \& {Furlanetto}, S.~R. 2008, \mnras, 384, 291

\bibitem[{{Ahn} {et~al.}(2012){Ahn}, {Iliev}, {Shapiro}, {Mellema}, {Koda}, \&
  {Mao}}]{2012ApJ...756L..16A}
{Ahn}, K., {Iliev}, I.~T., {Shapiro}, P.~R., {et~al.} 2012, \apjl, 756, L16

\bibitem[{{Ahn} {et~al.}(2015{\natexlab{a}}){Ahn}, {Mesinger}, {Alvarez}, \&
  {Chen}}]{2015aska.confE...3A}
{Ahn}, K., {Mesinger}, A., {Alvarez}, M.~A., \& {Chen}, X. 2015{\natexlab{a}},
  Advancing Astrophysics with the Square Kilometre Array (AASKA14), 3

\bibitem[{{Ahn} {et~al.}(2015{\natexlab{b}}){Ahn}, {Xu}, {Norman}, {Alvarez},
  \& {Wise}}]{2015ApJ...802....8A}
{Ahn}, K., {Xu}, H., {Norman}, M.~L., {Alvarez}, M.~A., \& {Wise}, J.~H.
  2015{\natexlab{b}}, \apj, 802, 8

\bibitem[{{Alvarez} {et~al.}(2006){Alvarez}, {Komatsu}, {Dor{\'e}}, \&
  {Shapiro}}]{2006ApJ...647..840A}
{Alvarez}, M.~A., {Komatsu}, E., {Dor{\'e}}, O., \& {Shapiro}, P.~R. 2006,
  \apj, 647, 840

\bibitem[{Asaba {et~al.}(2016)Asaba, Ichiki, \& Tashiro}]{PhysRevD.93.023518}
Asaba, S., Ichiki, K., \& Tashiro, H. 2016, Phys. Rev. D, 93, 023518

\bibitem[{{Baek} {et~al.}(2010){Baek}, {Semelin}, {Di Matteo}, {Revaz}, \&
  {Combes}}]{2010A&A...523A...4B}
{Baek}, S., {Semelin}, B., {Di Matteo}, P., {Revaz}, Y., \& {Combes}, F. 2010,
  \aap, 523, A4

\bibitem[{{Barger} {et~al.}(2009){Barger}, {Gao}, {Mao}, \&
  {Marfatia}}]{2009PhLB..673..173B}
{Barger}, V., {Gao}, Y., {Mao}, Y., \& {Marfatia}, D. 2009, Physics Letters B,
  673, 173

\bibitem[{{Bouwens} {et~al.}(2015){Bouwens}, {Illingworth}, {Oesch}, {Trenti},
  {Labb{\'e}}, {Bradley}, {Carollo}, {van Dokkum}, {Gonzalez}, {Holwerda},
  {Franx}, {Spitler}, {Smit}, \& {Magee}}]{2015ApJ...803...34B}
{Bouwens}, R.~J., {Illingworth}, G.~D., {Oesch}, P.~A., {et~al.} 2015, \apj,
  803, 34

\bibitem[{{Bromm} {et~al.}(2002){Bromm}, {Coppi}, \&
  {Larson}}]{2002ApJ...564...23B}
{Bromm}, V., {Coppi}, P.~S., \& {Larson}, R.~B. 2002, \apj, 564, 23

\bibitem[{{Carilli} {et~al.}(2002){Carilli}, {Gnedin}, \&
  {Owen}}]{2002ApJ...577...22C}
{Carilli}, C.~L., {Gnedin}, N.~Y., \& {Owen}, F. 2002, \apj, 577, 22

\bibitem[{{Chen} {et~al.}(2014){Chen}, {Heger}, {Woosley}, {Almgren}, \&
  {Whalen}}]{2014ApJ...792...44C}
{Chen}, K.-J., {Heger}, A., {Woosley}, S., {Almgren}, A., \& {Whalen}, D.~J.
  2014, \apj, 792, 44

\bibitem[{{Chen} \& {Miralda-Escud{\'e}}(2008)}]{2008ApJ...684...18C}
{Chen}, X. \& {Miralda-Escud{\'e}}, J. 2008, \apj, 684, 18

\bibitem[{{Choudhury} {et~al.}(2009){Choudhury}, {Haehnelt}, \&
  {Regan}}]{2009MNRAS.394..960C}
{Choudhury}, T.~R., {Haehnelt}, M.~G., \& {Regan}, J. 2009, \mnras, 394, 960

\bibitem[{{Ciardi} {et~al.}(2000){Ciardi}, {Ferrara}, {Governato}, \&
  {Jenkins}}]{2000MNRAS.314..611C}
{Ciardi}, B., {Ferrara}, A., {Governato}, F., \& {Jenkins}, A. 2000, \mnras,
  314, 611

\bibitem[{{Ciardi} {et~al.}(2015{\natexlab{a}}){Ciardi}, {Inoue}, {Abdalla},
  {Asad}, {Bernardi}, {Bolton}, {Brentjens}, {de Bruyn}, {Chapman}, {Daiboo},
  {Fernandez}, {Ghosh}, {Graziani}, {Harker}, {Iliev}, {Jeli{\'c}}, {Jensen},
  {Kazemi}, {Koopmans}, {Martinez}, {Maselli}, {Mellema}, {Offringa}, {Pandey},
  {Schaye}, {Thomas}, {Vedantham}, {Yatawatta}, \&
  {Zaroubi}}]{2015MNRAS.453..101C}
{Ciardi}, B., {Inoue}, S., {Abdalla}, F.~B., {et~al.} 2015{\natexlab{a}},
  \mnras, 453, 101

\bibitem[{{Ciardi} {et~al.}(2015{\natexlab{b}}){Ciardi}, {Inoue}, {Mack}, {Xu},
  \& {Bernardi}}]{2015aska.confE...6C}
{Ciardi}, B., {Inoue}, S., {Mack}, K., {Xu}, Y., \& {Bernardi}, G.
  2015{\natexlab{b}}, Advancing Astrophysics with the Square Kilometre Array
  (AASKA14), 6

\bibitem[{{Cooray}(2004)}]{2004PhRvD..70f3509C}
{Cooray}, A. 2004, \prd, 70, 063509

\bibitem[{{Dale} {et~al.}(2012){Dale}, {Ercolano}, \&
  {Bonnell}}]{2012MNRAS.424..377D}
{Dale}, J.~E., {Ercolano}, B., \& {Bonnell}, I.~A. 2012, \mnras, 424, 377

\bibitem[{{Evoli} {et~al.}(2014){Evoli}, {Mesinger}, \&
  {Ferrara}}]{2014JCAP...11..024E}
{Evoli}, C., {Mesinger}, A., \& {Ferrara}, A. 2014, JCAP, 11, 24

\bibitem[{{Fan} {et~al.}(2006){Fan}, {Carilli}, \&
  {Keating}}]{2006ARA&A..44..415F}
{Fan}, X., {Carilli}, C.~L., \& {Keating}, B. 2006, \araa, 44, 415

\bibitem[{{Fernandez} {et~al.}(2014){Fernandez}, {Zaroubi}, {Iliev}, {Mellema},
  \& {Jeli{\'c}}}]{2014MNRAS.440..298F}
{Fernandez}, E.~R., {Zaroubi}, S., {Iliev}, I.~T., {Mellema}, G., \&
  {Jeli{\'c}}, V. 2014, \mnras, 440, 298

\bibitem[{{Field}(1959)}]{1959ApJ...129..536F}
{Field}, G.~B. 1959, \apj, 129, 536

\bibitem[{{Furlanetto}(2006)}]{2006MNRAS.371..867F}
{Furlanetto}, S.~R. 2006, \mnras, 371, 867

\bibitem[{{Furlanetto} \& {Loeb}(2002)}]{2002ApJ...579....1F}
{Furlanetto}, S.~R. \& {Loeb}, A. 2002, \apj, 579, 1

\bibitem[{{Furlanetto} {et~al.}(2004){Furlanetto}, {Zaldarriaga}, \&
  {Hernquist}}]{2004ApJ...613...16F}
{Furlanetto}, S.~R., {Zaldarriaga}, M., \& {Hernquist}, L. 2004, \apj, 613, 16

\bibitem[{{Gardner} {et~al.}(2006){Gardner}, {Mather}, {Clampin}, {Doyon},
  {Greenhouse}, {Hammel}, {Hutchings}, {Jakobsen}, {Lilly}, {Long}, {Lunine},
  {McCaughrean}, {Mountain}, {Nella}, {Rieke}, {Rieke}, {Rix}, {Smith},
  {Sonneborn}, {Stiavelli}, {Stockman}, {Windhorst}, \&
  {Wright}}]{2006SSRv..123..485G}
{Gardner}, J.~P., {Mather}, J.~C., {Clampin}, M., {et~al.} 2006, \ssr, 123, 485

\bibitem[{{Giallongo} {et~al.}(2015){Giallongo}, {Grazian}, {Fiore}, {Fontana},
  {Pentericci}, {Vanzella}, {Dickinson}, {Kocevski}, {Castellano}, {Cristiani},
  {Ferguson}, {Finkelstein}, {Grogin}, {Hathi}, {Koekemoer}, {Newman}, \&
  {Salvato}}]{2015A&A...578A..83G}
{Giallongo}, E., {Grazian}, A., {Fiore}, F., {et~al.} 2015, \aap, 578, A83

\bibitem[{{Gnedin} \& {Ostriker}(1997)}]{1997ApJ...486..581G}
{Gnedin}, N.~Y. \& {Ostriker}, J.~P. 1997, \apj, 486, 581

\bibitem[{{Hasegawa} \& {Semelin}(2013)}]{2013MNRAS.428..154H}
{Hasegawa}, K. \& {Semelin}, B. 2013, \mnras, 428, 154

\bibitem[{{Hasegawa} {et~al.}(2009){Hasegawa}, {Umemura}, \&
  {Susa}}]{2009MNRAS.395.1280H}
{Hasegawa}, K., {Umemura}, M., \& {Susa}, H. 2009, \mnras, 395, 1280

\bibitem[{{Heger} \& {Woosley}(2010)}]{2010ApJ...724..341H}
{Heger}, A. \& {Woosley}, S.~E. 2010, \apj, 724, 341

\bibitem[{{Hirano} {et~al.}(2015){Hirano}, {Hosokawa}, {Yoshida}, {Omukai}, \&
  {Yorke}}]{2015MNRAS.448..568H}
{Hirano}, S., {Hosokawa}, T., {Yoshida}, N., {Omukai}, K., \& {Yorke}, H.~W.
  2015, \mnras, 448, 568

\bibitem[{{Hirano} {et~al.}(2014){Hirano}, {Hosokawa}, {Yoshida}, {Umeda},
  {Omukai}, {Chiaki}, \& {Yorke}}]{2014ApJ...781...60H}
{Hirano}, S., {Hosokawa}, T., {Yoshida}, N., {et~al.} 2014, \apj, 781, 60

\bibitem[{{Hosokawa} {et~al.}(2011){Hosokawa}, {Omukai}, {Yoshida}, \&
  {Yorke}}]{2011Sci...334.1250H}
{Hosokawa}, T., {Omukai}, K., {Yoshida}, N., \& {Yorke}, H.~W. 2011, Science,
  334, 1250

\bibitem[{{Iliev} {et~al.}(2014){Iliev}, {Mellema}, {Ahn}, {Shapiro}, {Mao}, \&
  {Pen}}]{2014MNRAS.439..725I}
{Iliev}, I.~T., {Mellema}, G., {Ahn}, K., {et~al.} 2014, \mnras, 439, 725

\bibitem[{{Iliev} {et~al.}(2006){Iliev}, {Mellema}, {Pen}, {Merz}, {Shapiro},
  \& {Alvarez}}]{2006MNRAS.369.1625I}
{Iliev}, I.~T., {Mellema}, G., {Pen}, U.-L., {et~al.} 2006, \mnras, 369, 1625

\bibitem[{{Jelic} {et~al.}(2015){Jelic}, {Ciardi}, {Fernandez}, {Tashiro}, \&
  {Vrbanec}}]{2015aska.confE...8J}
{Jelic}, V., {Ciardi}, B., {Fernandez}, E., {Tashiro}, H., \& {Vrbanec}, D.
  2015, Advancing Astrophysics with the Square Kilometre Array (AASKA14), 8

\bibitem[{{Kim} {et~al.}(2013{\natexlab{a}}){Kim}, {Power}, {Baugh}, {Wyithe},
  {Lacey}, {Lagos}, \& {Frenk}}]{2013MNRAS.428.3366K}
{Kim}, H.-S., {Power}, C., {Baugh}, C.~M., {et~al.} 2013{\natexlab{a}}, \mnras,
  428, 3366

\bibitem[{{Kim} {et~al.}(2013{\natexlab{b}}){Kim}, {Wyithe}, {Raskutti},
  {Lacey}, \& {Helly}}]{2013MNRAS.428.2467K}
{Kim}, H.-S., {Wyithe}, J.~S.~B., {Raskutti}, S., {Lacey}, C.~G., \& {Helly},
  J.~C. 2013{\natexlab{b}}, \mnras, 428, 2467

\bibitem[{{Kimm} \& {Cen}(2014)}]{2014ApJ...788..121K}
{Kimm}, T. \& {Cen}, R. 2014, \apj, 788, 121

\bibitem[{{Konno} {et~al.}(2014){Konno}, {Ouchi}, {Ono}, {Shimasaku},
  {Shibuya}, {Furusawa}, {Nakajima}, {Naito}, {Momose}, {Yuma}, \&
  {Iye}}]{2014ApJ...797...16K}
{Konno}, A., {Ouchi}, M., {Ono}, Y., {et~al.} 2014, \apj, 797, 16

\bibitem[{{Koopmans} {et~al.}(2015){Koopmans}, {Pritchard}, {Mellema},
  {Aguirre}, {Ahn}, {Barkana}, {van Bemmel}, {Bernardi}, {Bonaldi}, {Briggs},
  {de Bruyn}, {Chang}, {Chapman}, {Chen}, {Ciardi}, {Dayal}, {Ferrara},
  {Fialkov}, {Fiore}, {Ichiki}, {Illiev}, {Inoue}, {Jelic}, {Jones}, {Lazio},
  {Maio}, {Majumdar}, {Mack}, {Mesinger}, {Morales}, {Parsons}, {Pen},
  {Santos}, {Schneider}, {Semelin}, {de Souza}, {Subrahmanyan}, {Takeuchi},
  {Vedantham}, {Wagg}, {Webster}, {Wyithe}, {Datta}, \&
  {Trott}}]{2015aska.confE...1K}
{Koopmans}, L., {Pritchard}, J., {Mellema}, G., {et~al.} 2015, Advancing
  Astrophysics with the Square Kilometre Array (AASKA14), 1

\bibitem[{{Laureijs} {et~al.}(2011){Laureijs}, {Amiaux}, {Arduini},
  {Augu{\`e}res}, {Brinchmann}, {Cole}, {Cropper}, {Dabin}, {Duvet}, {Ealet},
  \& et~al.}]{2011arXiv1110.3193L}
{Laureijs}, R., {Amiaux}, J., {Arduini}, S., {et~al.} 2011, ArXiv e-prints
  1110.3193

\bibitem[{{Lidz} {et~al.}(2009){Lidz}, {Zahn}, {Furlanetto}, {McQuinn},
  {Hernquist}, \& {Zaldarriaga}}]{2009ApJ...690..252L}
{Lidz}, A., {Zahn}, O., {Furlanetto}, S.~R., {et~al.} 2009, \apj, 690, 252

\bibitem[{{Madau} {et~al.}(1997){Madau}, {Meiksin}, \&
  {Rees}}]{1997ApJ...475..429M}
{Madau}, P., {Meiksin}, A., \& {Rees}, M.~J. 1997, \apj, 475, 429

\bibitem[{{Maio} {et~al.}(2015){Maio}, {Ciardi}, \&
  {Koopmans}}]{2015aska.confE...9M}
{Maio}, U., {Ciardi}, B., \& {Koopmans}, L. 2015, Advancing Astrophysics with
  the Square Kilometre Array (AASKA14), 9

\bibitem[{{Maio} {et~al.}(2011){Maio}, {Koopmans}, \&
  {Ciardi}}]{2011MNRAS.412L..40M}
{Maio}, U., {Koopmans}, L.~V.~E., \& {Ciardi}, B. 2011, \mnras, 412, L40

\bibitem[{{Majumdar} {et~al.}(2014){Majumdar}, {Mellema}, {Datta}, {Jensen},
  {Choudhury}, {Bharadwaj}, \& {Friedrich}}]{2014MNRAS.443.2843M}
{Majumdar}, S., {Mellema}, G., {Datta}, K.~K., {et~al.} 2014, \mnras, 443, 2843

\bibitem[{{Mao} {et~al.}(2008){Mao}, {Tegmark}, {McQuinn}, {Zaldarriaga}, \&
  {Zahn}}]{2008PhRvD..78b3529M}
{Mao}, Y., {Tegmark}, M., {McQuinn}, M., {Zaldarriaga}, M., \& {Zahn}, O. 2008,
  \prd, 78, 023529

\bibitem[{{McQuinn} {et~al.}(2006){McQuinn}, {Zahn}, {Zaldarriaga},
  {Hernquist}, \& {Furlanetto}}]{2006ApJ...653..815M}
{McQuinn}, M., {Zahn}, O., {Zaldarriaga}, M., {Hernquist}, L., \& {Furlanetto},
  S.~R. 2006, \apj, 653, 815

\bibitem[{{Mellema} {et~al.}(2015){Mellema}, {Koopmans}, {Shukla}, {Datta},
  {Mesinger}, \& {Majumdar}}]{2015aska.confE..10M}
{Mellema}, G., {Koopmans}, L., {Shukla}, H., {et~al.} 2015, Advancing
  Astrophysics with the Square Kilometre Array (AASKA14), 10

\bibitem[{{Mesinger} {et~al.}(2014){Mesinger}, {Ewall-Wice}, \&
  {Hewitt}}]{2014MNRAS.439.3262M}
{Mesinger}, A., {Ewall-Wice}, A., \& {Hewitt}, J. 2014, \mnras, 439, 3262

\bibitem[{{Mesinger} {et~al.}(2015){Mesinger}, {Ferrara}, {Greig}, {Iliev},
  {Mellema}, {Pritchard}, \& {Santos}}]{2015aska.confE..11M}
{Mesinger}, A., {Ferrara}, A., {Greig}, B., {et~al.} 2015, Advancing
  Astrophysics with the Square Kilometre Array (AASKA14), 11

\bibitem[{{Mesinger} {et~al.}(2011){Mesinger}, {Furlanetto}, \&
  {Cen}}]{2011MNRAS.411..955M}
{Mesinger}, A., {Furlanetto}, S., \& {Cen}, R. 2011, \mnras, 411, 955

\bibitem[{{Moriya} {et~al.}(2010){Moriya}, {Yoshida}, {Tominaga}, {Blinnikov},
  {Maeda}, {Tanaka}, \& {Nomoto}}]{2010AIPC.1294..268M}
{Moriya}, T., {Yoshida}, N., {Tominaga}, N., {et~al.} 2010, in American
  Institute of Physics Conference Series, Vol. 1294, American Institute of
  Physics Conference Series, ed. D.~J. {Whalen}, V.~{Bromm}, \& N.~{Yoshida},
  268--269

\bibitem[{{Naoz} {et~al.}(2012){Naoz}, {Yoshida}, \&
  {Gnedin}}]{2012ApJ...747..128N}
{Naoz}, S., {Yoshida}, N., \& {Gnedin}, N.~Y. 2012, \apj, 747, 128

\bibitem[{{Naoz} {et~al.}(2013){Naoz}, {Yoshida}, \&
  {Gnedin}}]{2013ApJ...763...27N}
{Naoz}, S., {Yoshida}, N., \& {Gnedin}, N.~Y. 2013, \apj, 763, 27

\bibitem[{{Oesch} {et~al.}(2014){Oesch}, {Bouwens}, {Illingworth}, {Labb{\'e}},
  {Smit}, {Franx}, {van Dokkum}, {Momcheva}, {Ashby}, {Fazio}, {Huang},
  {Willner}, {Gonzalez}, {Magee}, {Trenti}, {Brammer}, {Skelton}, \&
  {Spitler}}]{2014ApJ...786..108O}
{Oesch}, P.~A., {Bouwens}, R.~J., {Illingworth}, G.~D., {et~al.} 2014, \apj,
  786, 108

\bibitem[{{Omukai} \& {Yoshii}(2003)}]{2003ApJ...599..746O}
{Omukai}, K. \& {Yoshii}, Y. 2003, \apj, 599, 746

\bibitem[{{O'Shea} \& {Norman}(2008)}]{2008ApJ...673...14O}
{O'Shea}, B.~W. \& {Norman}, M.~L. 2008, \apj, 673, 14

\bibitem[{{Ouchi} {et~al.}(2010){Ouchi}, {Shimasaku}, {Furusawa}, {Saito},
  {Yoshida}, {Akiyama}, {Ono}, {Yamada}, {Ota}, {Kashikawa}, {Iye}, {Kodama},
  {Okamura}, {Simpson}, \& {Yoshida}}]{2010ApJ...723..869O}
{Ouchi}, M., {Shimasaku}, K., {Furusawa}, H., {et~al.} 2010, \apj, 723, 869

\bibitem[{{Paardekooper} {et~al.}(2015){Paardekooper}, {Khochfar}, \& {Dalla
  Vecchia}}]{2015MNRAS.451.2544P}
{Paardekooper}, J.-P., {Khochfar}, S., \& {Dalla Vecchia}, C. 2015, \mnras,
  451, 2544

\bibitem[{{Pawlik} {et~al.}(2009){Pawlik}, {Schaye}, \& {van
  Scherpenzeel}}]{2009MNRAS.394.1812P}
{Pawlik}, A.~H., {Schaye}, J., \& {van Scherpenzeel}, E. 2009, \mnras, 394,
  1812

\bibitem[{{Petkova} \& {Maio}(2012)}]{2012MNRAS.422.3067P}
{Petkova}, M. \& {Maio}, U. 2012, \mnras, 422, 3067

\bibitem[{{Planck Collaboration} {et~al.}(2015){Planck Collaboration}, {Ade},
  {Aghanim}, {Arnaud}, {Ashdown}, {Aumont}, {Baccigalupi}, {Banday},
  {Barreiro}, {Bartlett}, \& et~al.}]{2015arXiv150201589P}
{Planck Collaboration}, {Ade}, P.~A.~R., {Aghanim}, N., {et~al.} 2015, ArXiv
  e-prints 1502.01589

\bibitem[{{Pritchard} {et~al.}(2015){Pritchard}, {Ichiki}, {Mesinger},
  {Metcalf}, {Pourtsidou}, {Santos}, {Abdalla}, {Chang}, {Chen}, {Weller}, \&
  {Zaroubi}}]{2015aska.confE..12P}
{Pritchard}, J., {Ichiki}, K., {Mesinger}, A., {et~al.} 2015, Advancing
  Astrophysics with the Square Kilometre Array (AASKA14), 12

\bibitem[{{Pritchard} \& {Pierpaoli}(2008)}]{2008PhRvD..78f5009P}
{Pritchard}, J.~R. \& {Pierpaoli}, E. 2008, \prd, 78, 065009

\bibitem[{{Razoumov} \& {Sommer-Larsen}(2010)}]{2010ApJ...710.1239R}
{Razoumov}, A.~O. \& {Sommer-Larsen}, J. 2010, \apj, 710, 1239

\bibitem[{{Ricotti} {et~al.}(2002){Ricotti}, {Gnedin}, \&
  {Shull}}]{2002ApJ...575...49R}
{Ricotti}, M., {Gnedin}, N.~Y., \& {Shull}, J.~M. 2002, \apj, 575, 49

\bibitem[{{Robertson} {et~al.}(2015){Robertson}, {Ellis}, {Furlanetto}, \&
  {Dunlop}}]{2015ApJ...802L..19R}
{Robertson}, B.~E., {Ellis}, R.~S., {Furlanetto}, S.~R., \& {Dunlop}, J.~S.
  2015, \apjl, 802, L19

\bibitem[{{Semelin} \& {Iliev}(2015)}]{2015aska.confE..13S}
{Semelin}, B. \& {Iliev}, I. 2015, Advancing Astrophysics with the Square
  Kilometre Array (AASKA14), 13

\bibitem[{{Sethi}(2005)}]{2005MNRAS.363..818S}
{Sethi}, S.~K. 2005, \mnras, 363, 818

\bibitem[{{Shimabukuro} {et~al.}(2014){Shimabukuro}, {Ichiki}, {Inoue}, \&
  {Yokoyama}}]{2014PhRvD..90h3003S}
{Shimabukuro}, H., {Ichiki}, K., {Inoue}, S., \& {Yokoyama}, S. 2014, \prd, 90,
  083003

\bibitem[{{Shimabukuro} {et~al.}(2015{\natexlab{a}}){Shimabukuro}, {Yoshiura},
  {Takahashi}, {Yokoyama}, \& {Ichiki}}]{2015arXiv150701335S}
{Shimabukuro}, H., {Yoshiura}, S., {Takahashi}, K., {Yokoyama}, S., \&
  {Ichiki}, K. 2015{\natexlab{a}}, ArXiv e-prints 1507.01335 Accepted in \mnras

\bibitem[{{Shimabukuro} {et~al.}(2015{\natexlab{b}}){Shimabukuro}, {Yoshiura},
  {Takahashi}, {Yokoyama}, \& {Ichiki}}]{2015MNRAS.451..467S}
{Shimabukuro}, H., {Yoshiura}, S., {Takahashi}, K., {Yokoyama}, S., \&
  {Ichiki}, K. 2015{\natexlab{b}}, \mnras, 451, 467

\bibitem[{{Sitwell} {et~al.}(2014){Sitwell}, {Mesinger}, {Ma}, \&
  {Sigurdson}}]{2014MNRAS.438.2664S}
{Sitwell}, M., {Mesinger}, A., {Ma}, Y.-Z., \& {Sigurdson}, K. 2014, \mnras,
  438, 2664

\bibitem[{{Slosar} {et~al.}(2007){Slosar}, {Cooray}, \&
  {Silk}}]{2007MNRAS.377..168S}
{Slosar}, A., {Cooray}, A., \& {Silk}, J.~I. 2007, \mnras, 377, 168

\bibitem[{{Smidt} {et~al.}(2015){Smidt}, {Whalen}, {Chatzopoulos}, {Wiggins},
  {Chen}, {Kozyreva}, \& {Even}}]{2015ApJ...805...44S}
{Smidt}, J., {Whalen}, D.~J., {Chatzopoulos}, E., {et~al.} 2015, \apj, 805, 44

\bibitem[{{Susa}(2007)}]{2007ApJ...659..908S}
{Susa}, H. 2007, \apj, 659, 908

\bibitem[{{Susa} {et~al.}(2014){Susa}, {Hasegawa}, \&
  {Tominaga}}]{2014ApJ...792...32S}
{Susa}, H., {Hasegawa}, K., \& {Tominaga}, N. 2014, \apj, 792, 32

\bibitem[{{Tanaka} {et~al.}(2012){Tanaka}, {Moriya}, {Yoshida}, \&
  {Nomoto}}]{2012MNRAS.422.2675T}
{Tanaka}, M., {Moriya}, T.~J., {Yoshida}, N., \& {Nomoto}, K. 2012, \mnras,
  422, 2675

\bibitem[{{Tashiro} {et~al.}(2010){Tashiro}, {Aghanim}, {Langer}, {Douspis},
  {Zaroubi}, \& {Jelic}}]{2010MNRAS.402.2617T}
{Tashiro}, H., {Aghanim}, N., {Langer}, M., {et~al.} 2010, \mnras, 402, 2617

\bibitem[{{Tassev} {et~al.}(2013){Tassev}, {Zaldarriaga}, \&
  {Eisenstein}}]{2013JCAP...06..036T}
{Tassev}, S., {Zaldarriaga}, M., \& {Eisenstein}, D.~J. 2013, JCAP, 6, 36

\bibitem[{{Toma} {et~al.}(2011){Toma}, {Sakamoto}, \&
  {M{\'e}sz{\'a}ros}}]{2011ApJ...731..127T}
{Toma}, K., {Sakamoto}, T., \& {M{\'e}sz{\'a}ros}, P. 2011, \apj, 731, 127

\bibitem[{{Trac} \& {Cen}(2007)}]{2007ApJ...671....1T}
{Trac}, H. \& {Cen}, R. 2007, \apj, 671, 1

\bibitem[{{Tseliakhovich} \& {Hirata}(2010)}]{2010PhRvD..82h3520T}
{Tseliakhovich}, D. \& {Hirata}, C. 2010, \prd, 82, 083520

\bibitem[{{Ueda} {et~al.}(2014){Ueda}, {Akiyama}, {Hasinger}, {Miyaji}, \&
  {Watson}}]{2014ApJ...786..104U}
{Ueda}, Y., {Akiyama}, M., {Hasinger}, G., {Miyaji}, T., \& {Watson}, M.~G.
  2014, \apj, 786, 104

\bibitem[{{Umemura} {et~al.}(2012){Umemura}, {Susa}, {Hasegawa}, {Suwa}, \&
  {Semelin}}]{2012PTEP.2012aA306U}
{Umemura}, M., {Susa}, H., {Hasegawa}, K., {Suwa}, T., \& {Semelin}, B. 2012,
  Progress of Theoretical and Experimental Physics, 2012, 010000

\bibitem[{{Viel} {et~al.}(2013){Viel}, {Becker}, {Bolton}, \&
  {Haehnelt}}]{2013PhRvD..88d3502V}
{Viel}, M., {Becker}, G.~D., {Bolton}, J.~S., \& {Haehnelt}, M.~G. 2013, \prd,
  88, 043502

\bibitem[{{Whalen} {et~al.}(2013){Whalen}, {Fryer}, {Holz}, {Heger}, {Woosley},
  {Stiavelli}, {Even}, \& {Frey}}]{2013ApJ...762L...6W}
{Whalen}, D.~J., {Fryer}, C.~L., {Holz}, D.~E., {et~al.} 2013, \apjl, 762, L6

\bibitem[{{Wiersma} {et~al.}(2013){Wiersma}, {Ciardi}, {Thomas}, {Harker},
  {Zaroubi}, {Bernardi}, {Brentjens}, {de Bruyn}, {Daiboo}, {Jelic}, {Kazemi},
  {Koopmans}, {Labropoulos}, {Martinez}, {Mellema}, {Offringa}, {Pandey},
  {Schaye}, {Veligatla}, {Vedantham}, \& {Yatawatta}}]{2013MNRAS.432.2615W}
{Wiersma}, R.~P.~C., {Ciardi}, B., {Thomas}, R.~M., {et~al.} 2013, \mnras, 432,
  2615

\bibitem[{{Wise} {et~al.}(2012){Wise}, {Abel}, {Turk}, {Norman}, \&
  {Smith}}]{2012MNRAS.427..311W}
{Wise}, J.~H., {Abel}, T., {Turk}, M.~J., {Norman}, M.~L., \& {Smith}, B.~D.
  2012, \mnras, 427, 311

\bibitem[{{Wise} \& {Cen}(2009)}]{2009ApJ...693..984W}
{Wise}, J.~H. \& {Cen}, R. 2009, \apj, 693, 984

\bibitem[{{Wise} {et~al.}(2014){Wise}, {Demchenko}, {Halicek}, {Norman},
  {Turk}, {Abel}, \& {Smith}}]{2014MNRAS.442.2560W}
{Wise}, J.~H., {Demchenko}, V.~G., {Halicek}, M.~T., {et~al.} 2014, \mnras,
  442, 2560

\bibitem[{{Worseck} {et~al.}(2014){Worseck}, {Prochaska}, {Hennawi}, \&
  {McQuinn}}]{2014arXiv1405.7405W}
{Worseck}, G., {Prochaska}, J.~X., {Hennawi}, J.~F., \& {McQuinn}, M. 2014,
  ArXiv e-prints 1405.7405

\bibitem[{{Wouthuysen}(1952)}]{1952AJ.....57R..31W}
{Wouthuysen}, S.~A. 1952, \aj, 57, 31

\bibitem[{{Wyithe} {et~al.}(2015){Wyithe}, {Geil}, \&
  {Kim}}]{2015aska.confE..15W}
{Wyithe}, S., {Geil}, P., \& {Kim}, H. 2015, Advancing Astrophysics with the
  Square Kilometre Array (AASKA14), 15

\bibitem[{{Xu} {et~al.}(2009){Xu}, {Chen}, {Fan}, {Trac}, \&
  {Cen}}]{2009ApJ...704.1396X}
{Xu}, Y., {Chen}, X., {Fan}, Z., {Trac}, H., \& {Cen}, R. 2009, \apj, 704, 1396

\bibitem[{{Yajima} {et~al.}(2011){Yajima}, {Choi}, \&
  {Nagamine}}]{2011MNRAS.412..411Y}
{Yajima}, H., {Choi}, J.-H., \& {Nagamine}, K. 2011, \mnras, 412, 411

\bibitem[{{Yajima} \& {Li}(2014)}]{2014MNRAS.445.3674Y}
{Yajima}, H. \& {Li}, Y. 2014, \mnras, 445, 3674

\bibitem[{{Yoshida} {et~al.}(2008){Yoshida}, {Omukai}, \&
  {Hernquist}}]{2008Sci...321..669Y}
{Yoshida}, N., {Omukai}, K., \& {Hernquist}, L. 2008, Science, 321, 669

\bibitem[{{Yoshida} {et~al.}(2006){Yoshida}, {Omukai}, {Hernquist}, \&
  {Abel}}]{2006ApJ...652....6Y}
{Yoshida}, N., {Omukai}, K., {Hernquist}, L., \& {Abel}, T. 2006, \apj, 652, 6

\bibitem[{{Yoshiura} {et~al.}(2016){Yoshiura}, {Hasegawa}, {Ichiki}, {Tashiro},
  {Shimabukuro}, \& {Takahashi}}]{2016arXiv160204407Y}
{Yoshiura}, S., {Hasegawa}, K., {Ichiki}, K., {et~al.} 2016, ArXiv 1602.04407

\bibitem[{{Yoshiura} {et~al.}(2015){Yoshiura}, {Shimabukuro}, {Takahashi},
  {Momose}, {Nakanishi}, \& {Imai}}]{2015MNRAS.451..266Y}
{Yoshiura}, S., {Shimabukuro}, H., {Takahashi}, K., {et~al.} 2015, \mnras, 451,
  266

\bibitem[{{Zahn} {et~al.}(2007){Zahn}, {Lidz}, {McQuinn}, {Dutta}, {Hernquist},
  {Zaldarriaga}, \& {Furlanetto}}]{2007ApJ...654...12Z}
{Zahn}, O., {Lidz}, A., {McQuinn}, M., {et~al.} 2007, \apj, 654, 12

\end{thebibliography}


\end{document}